\documentclass[12pt,aps,preprint,showpacs]{revtex4}

\usepackage{amsmath}
\usepackage{graphicx}

\begin{document}

\bibliographystyle{prsty}

\author{Matteo Nicoli}
\affiliation{Grupo Interdisciplinar de Sistemas Complejos (GISC), Departamento de Matem\'aticas,
Universidad Carlos III de Madrid, Avenida de la Universidad 30, 28911 Legan\'{e}s, Spain}

\author{Mario Castro}
\affiliation{Grupo de Din\'amica No Lineal and GISC, Escuela T\'ecnica Superior de Ingenier\'{\i}a (ICAI)\\
Universidad Pontificia Comillas, E-28015 Madrid, Spain}

\author{Rodolfo Cuerno}
\affiliation{Departamento de Matem\'aticas and GISC, Universidad
Carlos III de Madrid, Avenida de la Universidad 30, 28911 Legan\'{e}s, Spain}

\title{Unified moving boundary model with fluctuations for unstable diffusive growth}

\begin{abstract}
We study a moving boundary model of non-conserved interface growth that implements the
interplay between diffusive matter transport and aggregation kinetics at the interface.
Conspicuous examples are found in thin film production by chemical vapor deposition and
electrochemical deposition. The model also incorporates noise terms that account for
fluctuations in the diffusive and in the attachment processes. A small slope approximation
allows us to derive effective interface evolution equations (IEE) in which parameters are
related to those of the full moving boundary problem. In particular, the form of the linear
dispersion relation of the IEE changes drastically for slow or for instantaneous attachment
kinetics. In the former case the IEE takes the form of the well-known (noisy)
Kuramoto-Sivashinsky equation, showing a morphological instability at short times that evolves
into kinetic roughening of the Kardar-Parisi-Zhang class. In the instantaneous kinetics limit,
the IEE combines
Mullins-Sekerka linear dispersion relation with a KPZ nonlinearity, and we provide a
numerical study of the ensuing dynamics. In all cases, the long preasymptotic transients can
account for the experimental difficulties to observe KPZ scaling. We also compare our results
with relevant data from experiments and discrete models.
\end{abstract}


\maketitle

\section{Introduction}

Ever since the beginning of the nineteenth century~\cite{Rosebrugh,Agar,Ibl},
diffusion-limited growth has attracted the attention of physicists, due to its experimental
ubiquity and, partly, because it is amenable to continuum descriptions that are sometimes
solvable. For instance, electrochemical deposition (ECD) of metals \cite{ecd,Bard} has been
and still is~\cite{lafouresse,kkrug,osafo} a subject of intense study during this time due
to its (in principle) experimental simplicity and its many technological applications:
A deposit grows on the cathode when a potential difference is set between two metallic
electrodes in a salt solution (generally of Cu, Ag or Zn). Another interesting system of
a conceptually similar type is chemical vapor deposition (CVD) \cite{cvd}, in which a
deposit grows from a vapor phase through the incorporation of a reacting species which
attaches via chemical reactions once it reaches the aggregate. CVD is one of the techniques
of choice for the fabrication of many microelectronic devices, and is currently being the
object of intense study~\cite{pelliccione2006bds, white2006cdr, yanguasgil2006iad}, partly
motivated by its use also in emerging fields of Science and Technology, such as
Microfluidics~\cite{tabeling:2005}. The practical relevance of ECD and CVD~\cite{ecd,Bard,cvd}
perhaps makes them appear as two paradigmatic examples of a larger class of growth systems
(that will be referred to henceforth as {\em diffusive}) in which dynamics is a result of
the competition between diffusive transport and attachment kinetics at the aggregate interface.
Given that growth dynamics in these processes is not constrained in principle by mass
conservation, they provide important examples of non-conserved growth \cite{Barabasi}.

Despite the great amount of work devoted to these systems, they still pose important challenges
to the detailed understanding of the very different structures grown under diverse conditions,
whose geometries range from fractal to columnar. Partial progress has been achieved so far
through the study of the time evolution of the aggregate surface and its
roughness~\cite{Barabasi,cuerno:2004}. In particular, a very successful theoretical framework
for such type of study has been the use of stochastic growth equations for the interface height.
Thus, $e.g.$ the celebrated Kardar-Parisi-Zhang (KPZ) equation~\cite{KPZ},
\begin{equation}
\frac{\partial h}{\partial t} = V + \nu \nabla^2 h +
\frac{V}{2} (\nabla h)^2 + \eta(\mathbf{r},t) ,
\label{kpzeq}
\end{equation}
has been postulated as a universal model of non-conserved rough interface growth.
In \eqref{kpzeq}, $\nu$ is a {\em positive} constant, $\eta(\mathbf{r},t)$ is an uncorrelated
Gaussian noise representing fluctuations, $e.g.$, in a flux of depositing particles, and $V$
is the average surface growth velocity.
Many times the KPZ equation has been put forward as the description of specific experimental
growth systems based on symmetry considerations and universality arguments. In view of the
fact \cite{cuerno:2007} that very few experiments have been reported which are compatible
with the predictions of the KPZ equation (two examples in ECD and CVD are provided in
\cite{Schilardi,Ojeda}), the main drawback of such a theoretical approach \cite{cuerno:2007}
is that, in most cases, this coarse-grained level of description does not allow to make a
connection between the experimental and the theoretical parameters, so that it is difficult
to assess the cause of the disagreement between theoretical description and experimental
observations. Moreover, such an approach does not allow to include in a systematic way other
potentially important physical mechanisms, such as $e.g.$ non-local effects (typical of
diffusive systems) or fluctuations related to mass transport within the dilute phase.

In previous work~\cite{ourprl,cuerno:2002} we have put forward an argument as to why many
experiments on non-conserved interface growth, such as by ECD and CVD, rarely reproduce the
KPZ roughness exponents. Namely, morphological instabilities usually occur in these and many
other growth systems that induce long crossovers making the asymptotic KPZ behavior hard to
observe. In this paper, we substantiate further such an approach to the problem of non-conserved
growth, by constructing a model that incorporates the main constitutive laws common to diffusive
growth systems, from which an effective stochastic growth equation can be explicitly derived from first principles.

The aim of this work is threefold: {\em First}, our basic (stochastic) moving boundary problem can
be explicitly related to realistic CVD and (simplified) ECD systems. We thus provide a novel
unified picture of these two growth techniques, that makes explicit their common features.
Nevertheless, due to the generality of the basic constitutive laws that we assume, we
expect our model to have implications also for different growth procedures that can
be described as {\em diffusive} in the sense described above. {\em Second},
we benefit from the model formulation in terms of constitutive laws in order to derive the
dependence of coefficients of the effective interface equation with physical parameters.
This result seems to be new in the context of diffusive growth, and will allow us to show that
the form (and properties) of the effective height equation depends crucially on the efficiency
of attachment kinetics. Specifically, if the kinetics at the surface is instantaneous, {\em i.e.},
if the particles aggregate with probability close to unity when they arrive at the surface, the
system can be described by a new equation which is morphologically unstable, but that still
provides non-KPZ scale invariance of the interface fluctuations at large scales.
This new result reinforces our previous conclusions \cite{ourprl,cuerno:2002}, on the
experimental irrelevance of KPZ scaling in diffusive growth systems.
On the other hand, for slow interface kinetics the effective evolution equation is the well-known
(stochastic) Kuramoto-Sivashinsky equation \cite{kuramoto}, that displays a qualitatively similar
dynamics, albeit with a long scale behavior that does fall into the KPZ class
\cite{cuerno:1995,cuerno:1995b,Ueno:Kuramoto}. {\em Third}, we will interpret some results
from experiments and discrete models under the light of our continuum theory both qualitatively and quantitatively.
As (the deterministic limit of) our model has been profusely tested in the case of CVD, we will
mostly consider experiments and models from the ECD context.

The paper is organized as follows. We describe in Sec.\ II our moving boundary formulation of
diffusive growth systems, including noise terms related to fluctuations in diffusive currents and
relaxation events. Section III reports a linear stability analysis of the ensuing unified model of
ECD and CVD. Using a small slope approximation, we derive in Section IV a universal nonlinear
stochastic equation for the aggregate surface that is numerically studied in the novel case of
infinitely fast kinetics. Sec.\ V is devoted to making a connection with several experiments and
discrete models on diffusive growth systems. Finally, we conclude in Sec.\ VI with a discussion
of our results, which will allow us to suggest a reasonably approximate picture of non-conserved
growth. Some technical details are given in the appendices.

\section{Moving boundary problem}
As it turns out, our description of diffusive growth systems takes a form whose deterministic limit
has been long studied in the context of CVD. Thus, we first review the classic constitutive equations
of CVD~\cite{Jansen,Palmer88,Bales}, and subsequently consider the effect of noise due to the
fluctuations related to the different relaxation mechanisms involved. Then, we write the equations
of ECD growth in a form that unifies this technique with CVD.

\subsection{Chemical vapor deposition}
\label{cvd_sec}

\begin{figure}[!ht]
\includegraphics[width=\textwidth]{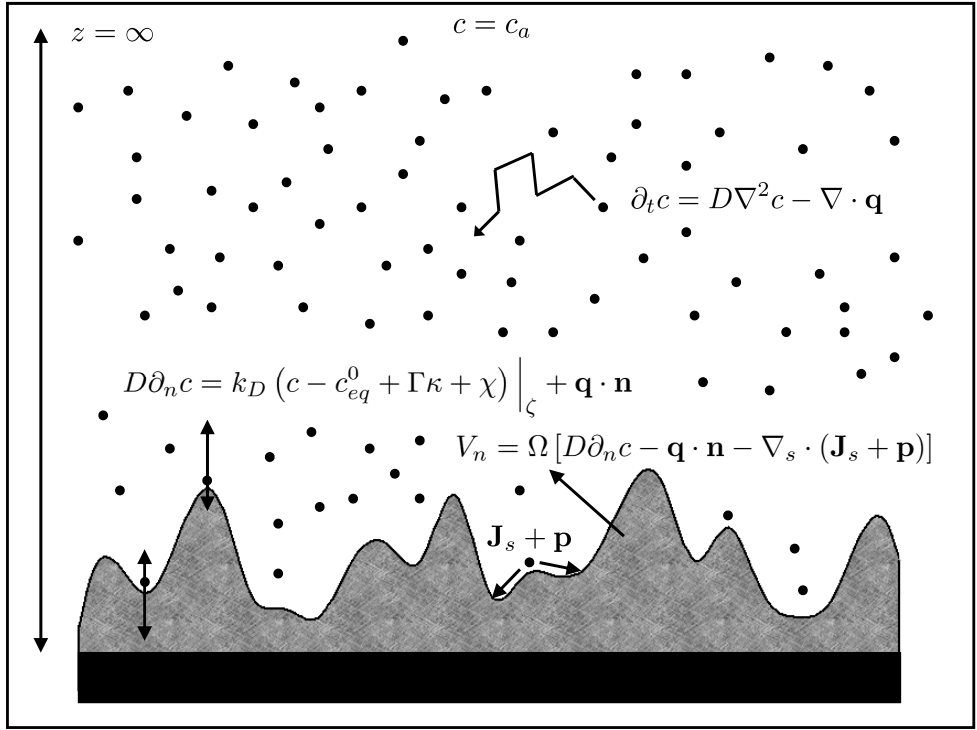}
\caption{Schematic representation of a model CVD growth system. Black points represent aggregating
units diffusing in the dilute phase. The different transport mechanisms (bulk diffusion, attachment and
surface diffusion) are indicated in the figure by their corresponding equations. For the definitions of the
noise terms $(\mathbf{q},\mathbf{p},\chi)$ see Sec. IIC.}
\label{esquemaCVD}
\end{figure}

A stagnant diffusion layer of infinite vertical extent is assumed to exist
above the substrate upon which an aggregate will grow, see a sketch in Fig.\ \ref{esquemaCVD}.
This approach implies
that the length of the stagnant layer (typically of the order of cm) is much larger than the typical
thickness of the deposit (in the range of microns). The particles within the vapor diffuse randomly
until they arrive at the surface, react and aggregate to it. The concentration of these particles,
$c(x,z,t)\equiv c({\bf r},t)$, obeys the diffusion equation
\begin{equation}
\partial_tc=D\nabla^2c.
\label{dif_eq_cvd}
\end{equation}
In the experiments, the mean concentration at the top of the stagnant layer is chosen to be a
constant, equal to the initial average concentration $c_a$.

Besides this, mass is conserved at the aggregate surface, so the local normal
velocity at an arbitrary point on the surface is given by
\begin{equation}
V_n=\Omega D {\nabla} c\cdot {\bf n}-\Omega {\nabla}_s\cdot {\bf J}_s,
\label{velo_cvd}
\end{equation}
where $\Omega$ is the molar volume of the aggregate and ${\bf n}$ is the local unit normal,
exterior to the aggregate. The last equation expresses the fact that growth takes place along
the local normal direction (usually referred to as {\em conformal growth} in the CVD literature)
and is due to the arrival of particles from the vapor [the first term in Eq.\ (\ref{velo_cvd})] and
via surface diffusion (${\bf J}_s$ stands for the diffusing particle current over the aggregate
surface and ${\nabla}_s$ is the surface gradient).

Moreover, the particle concentration, $c$, and its gradient at the surface are related through
the mixed boundary condition
\begin{equation}
k_D(c-c^0_{eq}+\Gamma\kappa)\Big|_{\zeta(x,t)}=D{\nabla} c\cdot{\bf n}\Big|_{\zeta(x,t)},
\label{mixta_cvd}
\end{equation}
where $c^0_{eq}$ is the local equilibrium concentration of a flat interface in contact with its
vapor, and $\zeta(x,t)$ is the local surface height. This equation is closely related to the
probability of a particle to stick to the surface when it reaches it (see below).

In summary, Eqs.\ (\ref{dif_eq_cvd}) and (\ref{velo_cvd}) describe diffusive
transport in the vapor phase and the way that particles attach to the
growing aggregate. Let us concentrate on the physical meaning of
Eq.\ (\ref{mixta_cvd}). The parameter $\Gamma$ is related to temperature
\cite{mullins:1957,mullins:1959} as
$\Gamma=\gamma c^0_{eq}\Omega/(k_BT)$,
where $\gamma$ is the surface tension (that will be assumed a constant) and
$\kappa=(\partial_{xx}\zeta) \left[1+(\partial_x\zeta)^2\right]^{-3/2}$ is the surface curvature.
The boundary condition (\ref{mixta_cvd}) can be obtained analytically $e.g.$ from kinetic
theory by computing the probability distribution for a random walker close to a partially
absorbing boundary. There, the particles have a sticking probability, $s$,
of aggregating irreversibly ({\em i.e.}, attachment is not deterministic). In such a case~\cite{Naqvi}
\begin{equation}
k_D=\frac{s}{2-s}DL^{-1}_{\rm mfp} ,
\label{k_D_sticking}
\end{equation}
where $L_{\rm mfp}$ is the particle mean free path. Assuming that
$L_{\rm mfp}$ is sufficiently small we find two limits in
Eq.\ (\ref{k_D_sticking}):  If the sticking probability vanishes ($s=0$)
then $\nabla c=0$ at the boundary, so the aggregate
does not grow. On the contrary, if the sticking probability is close to
unity (provided $L_{\rm mfp}$ is small enough), then $k_D$ takes very
large values and equation (\ref{mixta_cvd}) reduces to the well-known
Gibbs-Thomson relation~\cite{mullins:1957,mullins:1959}, which incorporates into
the equations the fact that concentration is different in regions with different curvature.
In summary, Eq.\ (\ref{mixta_cvd}) gives a simple macroscopic interpretation of a microscopic
parameter, the sticking probability, and allows to quantify the efficiency of the chemical
reactions leading to species attachment at the interface.

\subsection{Electrochemical deposition}
\label{ecd_sec}

\begin{figure}[!ht]
\includegraphics[width=\textwidth]{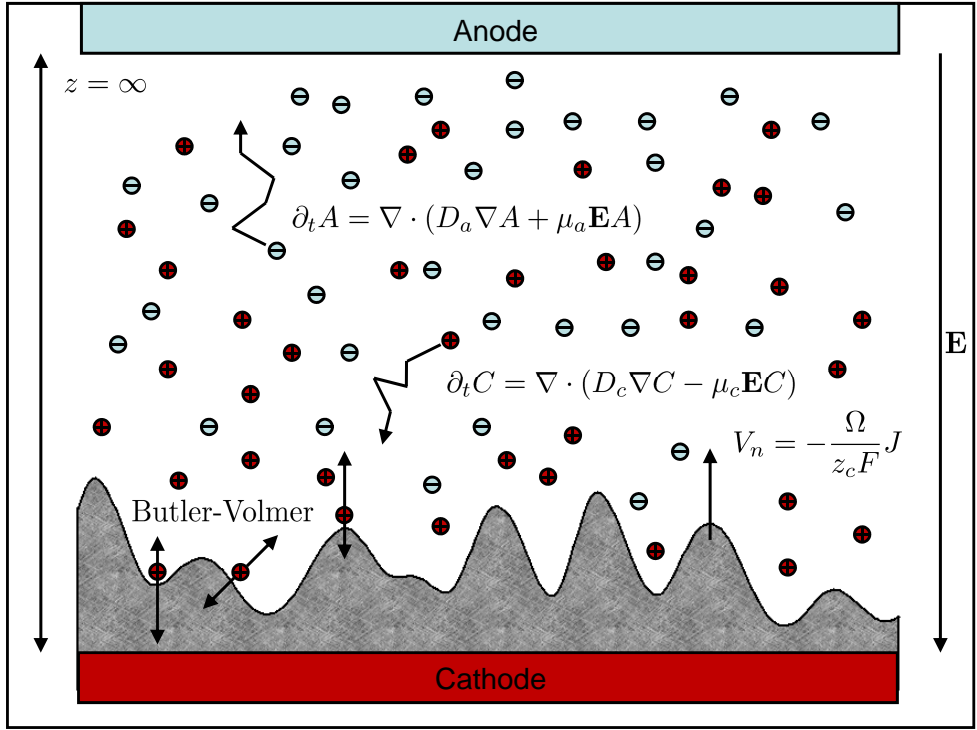}
\caption{(color online) Schematic representation of a model ECD growth system. Cations
migrate towards the cathode (lower side) while anions migrate towards the anode
(upper side) in an infinite cell. The different transport mechanisms (bulk diffusion and drift,
cation reduction) are indicated in the figure by their corresponding equations.}
\label{esquemaECD}
\end{figure}

In an electrochemical experiment, dynamics is more complex than in the CVD system as
represented above, due to the existence of two different species subject to transport
(anions and cations)~\cite{noteaboutspecies} and an imposed electric field.
For a visual reference, see Fig.\ \ref{esquemaECD}. Although more elaborate treatments
of these can be performed \cite{sundstroem:1995,haataja:2003a,haataja:2003b}, qualitatively
the morphological results are similar to the more simplified description we will be making in
what follows. The virtues of the latter include an explicit mapping to the CVD system and
explicit experimental verification.

Thus, in ECD, mass transport is not only due to diffusion in the dilute phase, but also due
to electromigration and convection.  Let $C$ and $A$ be the
concentration of cations and anions, respectively, then
\begin{eqnarray}
&\partial_t C=-\nabla\cdot{\bf J}_c,&=
	\nabla\cdot\left(D_c\nabla C-\mu_c{\bf E}C-{\bf v}C\right),
\label{difusion_C}\\
&\partial_t A=-\nabla\cdot{\bf J}_a,&=
	\nabla\cdot\left(D_a\nabla A+\mu_a{\bf E}A-{\bf v}A\right),
\label{difusion_A}
\end{eqnarray}
where $D_{c,a}$ are, respectively, the cationic and anionic diffusion
coefficients, $\mu_{c,a}$ are their mobilities, and $\bf E$ is the electric field
through the cell, which obeys the Poisson equation,
\begin{equation}
\nabla\cdot{\bf E}=-\nabla^2\phi=e N_A(z_cC-z_aA)/\varepsilon,
\label{poisson_couple}
\end{equation}
with
$N_A$ the Avogadro constant,
$ez_c$ and $-ez_a$ being the cationic and anionic charges, respectively,
$\phi$ the electric potential, and $\varepsilon$ the fluid permittivity.
The velocity  $\bf v$ of the fluid obeys the Navier-Stokes equation,
although we will assume this velocity to vanish in very thin
cells~\cite{Fleury}.

Another interesting experimental variable is the electric current density, $J$, given by
\begin{equation}
J=F(z_c\mu_cC+z_a\mu_aA)E=\sigma(t) E,
\end{equation}
where $\sigma(t)$ is the apparent electric conductivity.
Many experiments exploit the capability of tuning several parameters while
maintaining $J$ constant (galvanostatic conditions), hence the relevance
of this parameter.

In order to better understand the way in which the particles evolve in the cell,
we need to follow their dynamics when the external electric field is switched on.
Thus, the cation and anion concentrations are initially constant and uniform
across the cell and then, once the electric field is applied,
anions move towards the anode and cations move towards the cathode.
Cations reduce at the cathode thus forming an aggregate of neutral
particles.  On the contrary, anions do not aggregate, rather, they merely pile up at the
anode, which is dissolving at the same rate as the cations aggregate in the
cathode. Hence, the number of cations remains a constant.

Mathematically, this mechanism of aggregation can be expressed as a boundary
condition for the cation concentration. Before introducing such a boundary
condition we will simplify the set of diffusion equations
(\ref{difusion_C})-(\ref{difusion_A}) following
Refs.\ \cite{CommentChazalviel,Leger2}.
Let us consider that the deposit moves with a given constant velocity $V$,
in such a way that, in the frame of reference co-moving with the surface, $z=0$
is the position of the mean height and $z\rightarrow\infty$ represents the position
of the anode (thus, we are dealing with the case in which the height of the
aggregate is negligible with respect to the electrode separation).
Moreover, we will assume that the system is under galvanostatic conditions, namely,
that the current density at the cathode, $J$, is maintained constant. Thus, the problem
can be separated into two spatial regions: far enough from and close to the cathode.

At distances larger than the typical diffusion length, $l_D=D/V$, the net
charge is zero, so that $z_aA=z_cC$. Hence, multiplying Eq.\ (\ref{difusion_C}) by
$z_c\mu_a$, and adding Eq.\ (\ref{difusion_A}) multiplied by $z_a\mu_c$
we have
\begin{equation}
\partial_tC=D\nabla^2C,\quad \partial_t A=D\nabla^2 A.
\end{equation}
In both equations we have used the ambipolar diffusion coefficient,
given by
\begin{equation}
D=\frac{\mu_cD_a+\mu_aD_c}{\mu_a+\mu_c}.
\end{equation}
Hence, mass transport reduces to a single-variable diffusion equation.
It is also important that the electroneutrality condition ($z_aA=z_cC$)
implies that the mean interface velocity is equal to the anion migration
velocity, that is, $V=\mu_aE_\infty$, where $E_\infty$ is the
electric field very far from the cathode  (see
Refs.\ \cite{CommentChazalviel,Leger2} for further details), and then
\begin{equation}
\frac{J}{F}=z_c{\bf J}_c-z_a{\bf J}_a=-\frac{VF z_c \,C_a}{1-t_c},
\end{equation}
where
$t_c=\mu_c/(\mu_a+\mu_c)$,
and $C_a$ is the initial cation concentration. Finally, we must provide an
equation to describe cation attachment. As a point of departure we will
take a relation between the charge transport through the interface and
its local properties, given by the well-known Butler-Volmer (BV)
equation~\cite{ecd,Bard,sundstroem:1995,Kashchiev}
\begin{equation}
J=J_0\left[ e^{(1-b)\eta z_cF / RT}-
e^{-(b \eta + \eta_s) z_cF / RT} C_{\zeta}/C_a \right],
\label{Butler_Volmer}
\end{equation}
where $J_0$ is the exchange current density in equilibrium,
$b$ is a coefficient which ranges from $0$ to $1$, and estimates
the asymmetry of the energy barrier related to the cation reduction reaction, and
$\eta=\Delta \phi-\Delta \phi_{eq}$
is the overpotential, from which a surface curvature contribution $\eta_s$ has
been singled out, of the form
\begin{equation}
\frac{z_cF\eta_s}{RT}=\frac{\Omega\gamma}{RT}\kappa=\Gamma\Omega\kappa,
\end{equation}
where we have defined the parameter $\Gamma=\gamma/RT$ in the ECD context.
The first term on the right hand side of Eq.\ (\ref{Butler_Volmer})
is proportional to the rate of the backward reaction,
$X\rightarrow X^{n+}+ne^-$, and the second one is proportional to the rate of the
forward reaction, $X^{n+}+ne^-\rightarrow X$.
The factor $C_\zeta$ (the concentration at the surface) is due to the supply
of cations at the surface. Since the flux of anions through the cathode is zero
(because they neither react nor aggregate) the electric current density at the
aggregate surface is only due to the cations, and the charge current
is proportional to the cation current. Hence~\cite{ecd,Leger2000},
\begin{equation}
J=-\frac{z_cD_cFV}{1-t_c}\,\nabla C\cdot{\bf n}\Big|_\zeta.
\label{V_J}
\end{equation}
This equation, combined with Eq.\ (\ref{Butler_Volmer}), provides
a mixed boundary condition which relates the cation concentration at the
boundary with its gradient.
In order to cast it into a shape that recalls the CVD relation, we define
\begin{eqnarray}
K_D &=& \frac{J_0}{z_cFC_a}e^{-bz_cF\eta/RT},\\
C^0_{eq}&=& C_ae^{z_cF\eta/RT}, \label{c0eq_ca}
\end{eqnarray}
and obtain from \eqref{Butler_Volmer}, \eqref{V_J}
\begin{equation}
\frac{D_c}{1-t_c}\nabla C\cdot{\bf n}\Big|_\zeta=K_D(C-C^0_{eq})\Big|_\zeta.
\label{elastica}
\end{equation}
The coefficient $K_D$ is related with the sticking probability for cations:
if the aggregation is very effective (large sticking probability) the overpotential
is a large negative quantity and then $K_D$ grows exponentially. In addition, the
concentration $C^0_{eq}$ decreases. Hence, we can approximate Eq.\ (\ref{elastica}) by
$C\simeq C^0_{eq}$.  In the limit when every particle which arrives at the surface
sticks irreversibly, the solution cannot supply enough particles, $C=0$, and the
current density takes its maximum value. This value of the current is called limiting
current density. On the other hand, if the sticking  probability is small, the system
is always close to equilibrium and $\nabla C\simeq 0$, so that the net current
is zero.

Finally, we close the system with an equation for mass conservation
at the boundary. Note that the local velocity of the aggregate surface
is proportional to the flux of particles arriving to it, therefore
\begin{equation}
V_n=-\Omega{\bf J}_c\cdot{\bf n}=-\frac{\Omega}{z_cF}J,
\end{equation}
where $\Omega$ is the molar volume, here defined as the ratio of the metal
molar mass, $M$, and the aggregate mean density, $\rho$.
For a flat front, $V_n=V$, hence comparing this equation with
Eq.\ (\ref{V_J}) we find
\begin{equation}
\Omega=\frac{M}{\rho}=\frac{1-t_c}{C_a},
\label{Omega_c_a}
\end{equation}
a relationship which has been previously proposed theoretically~\cite{Leger2000}
and experimentally verified~\cite{Fleury}, thus supporting the hypotheses made in this section.

Surely, the reader has noticed that the last equations resemble those
for CVD. To emphasize this similarity, we define new variables and parameters as
\begin{equation}
 c = R_c C, \quad c_a= R_c C_a,\quad c^0_{eq}=R_c C^0_{eq},
 \quad k_D=\frac{K_D}{R_c},
\end{equation}
where $R_c \equiv D_c/[D(1-t_c)]$.
With these definitions, Eqs. (\ref{dif_eq_cvd})-(\ref{mixta_cvd})
describe (under the physical assumptions made above) the evolution of
both diffusive growth systems, CVD and ECD.

\subsection{The role of fluctuations}

The set of equations presented in the previous section describes the evolution
of the mean value of the concentration so that, formally, we can track the
position of the interface at any instant. However, it explicitly ignores the (thermal)
fluctuations related to the different transport and relaxation mechanisms involved.
In order to account for these, we define the stochastic functions ${\bf q}$, ${\bf p}$
and $\chi$ as the fluctuations in the flux of particles in the dilute phase ($-D\nabla c$),
in the surface-diffusing particle current (${\bf J}_s$), and in the equilibrium
concentration value at the interface, respectively. We choose these noise terms,
${\bf q}$, ${\bf p}$ and $\chi$, to have zero mean value and correlations given by
\begin{eqnarray}
\langle q_i({\bf r},t)\, q_j({\bf r}^\prime,t^\prime)\rangle&=&Q\, \delta_{ij}
\delta({\bf r}-{\bf r}^\prime)\delta(t-t^\prime),\label{Q_def} \\
\langle p_i({\bf r},t)\, p_j({\bf r}^\prime,t^\prime)\rangle&=&P\, \delta_{ij}
\frac{\delta({\bf r}-{\bf r}^\prime)\delta(t-t^\prime)}{\sqrt{1+(\partial_x\zeta)^2}},\label{P_def} \\
\langle \chi({\bf r},t)\, \chi({\bf r}^\prime,t^\prime)\rangle&=&I \,
\frac{\delta({\bf r}-{\bf
r}^\prime)\delta(t-t^\prime)}{\sqrt{1+(\partial_x\zeta)^2}},\label{I_def}
\end{eqnarray}
where $i,j$ denote vector components and $Q$, $P$, and $I$ will be determined from
the equilibrium fluctuations following \cite{Karma,Pierre}. Finally, the factor
$\sqrt{1+(\partial_x\zeta)^2}$ in \eqref{P_def}, \eqref{I_def} ensures that the noise
strength is independent of the surface orientation.

Thus, the stochastic moving boundary problem we propose to describe diffusive growth
has the form
\begin{eqnarray}
\partial_tc&=&
D\nabla^2c
-\nabla\cdot{\bf q},\label{n_difusion_eq}\\
D\partial_nc &=&k_D(c-c^0_{eq}+\Gamma\kappa+\chi)\Big|_\zeta+
{\bf q}\cdot{\bf n},\label{n_boundary_cond}\\
V_n&=&\Omega \, \Big[ D\partial_nc-\nabla_s\cdot{\bf J}_s
-{\bf q}\cdot{\bf n} -\nabla_s\cdot{\bf p}\Big], \label{n_velocidad} \\
\lim_{z\rightarrow\infty} & & \hskip -0.5 truecm  c(x,z;t) =c_a.\label{C_anode}
\end{eqnarray}
In \eqref{n_velocidad} the surface diffusion term, $\nabla_s\cdot{\bf J}_s$, is proportional
to the surface diffusion coefficient $D_s$ and the surface concentration of particles
$\nu_s$; moreover, this term is related to the local surface curvature \cite{mullins:1957,mullins:1959}
\begin{equation}
\nabla_s\cdot{\bf J}_s=B\nabla_s^2\kappa \label{B_Mullins}.
\end{equation}
In order to determine the values of the coefficients $Q$, $P$ and $I$ defined in equations
(\ref{Q_def})-(\ref{I_def}), we use a local equilibrium hypothesis \cite{Landau,Pierre}.
To begin with, let us consider an ideal concentration $c_a$ of
randomly distributed particles. The probability of finding $n$ particles in a given volume
is given by a Poisson distribution. The mean and variance of this distribution are $c_a$,
hence the concentration $c$ satisfies
\begin{equation}
\langle (c({\bf r},t)-c_a)(c({\bf r}^\prime,t^\prime)-c_a)\rangle=
c_a\, \delta({\bf r}-{\bf r}^\prime)\, \delta(t-t^\prime).
\label{corr_eq}
\end{equation}
This equation will allow us to determine $Q$. First,
we write Eq.\ (\ref{n_difusion_eq}) as
\begin{equation}
\partial_t(c-c_a)=D\nabla^2(c-c_a)-\nabla\cdot{\bf q}.
\label{shifted_dif_eq}
\end{equation}
Let $c_{{\bf k}\omega}$ and ${\bf q}_{{\bf k}\omega}$  be the
Fourier transforms of  $[c({\bf r},t)-c_a]$ and  $\bf q$, respectively,
\begin{eqnarray}
c_{{\bf k}\omega}&=& \int dt \,
e^{-i\omega t}\int d{\bf r} \,e^{-i{\bf k}\cdot{\bf r}} [c({\bf
r},t)-c_a],\\
{\bf q}_{{\bf k}\omega}&=& \int
dt\, e^{-i\omega t}\int d{\bf r} \,e^{-i{\bf k}\cdot{\bf r}} {\bf q}({\bf
r},t).
\end{eqnarray}
Writing Eq.\ \eqref{shifted_dif_eq} in momentum-frequency space and comparing
with the Fourier transform of Eq.\ \eqref{corr_eq},
we find that the spectrum of equilibrium fluctuations is (after integrating out $\omega$)
\begin{equation}
\langle c_{\bf k}\, c_{\bf -k}\rangle=\frac{Q}{2D}.
\end{equation}
Hence, using (\ref{corr_eq}), we obtain $Q=2Dc_a$.

Similarly, in order to determine $I$, we just note that the equilibrium
distribution of a curved interface is given by the Boltzmann distribution
\begin{equation}
P(\{\zeta\})\sim\exp\left[-\frac{{\mathcal H}(\{\zeta\})}{k_BT} \right],
\label{Prob_boltz}
\end{equation}
$\mathcal H$ being a functional that measures the amount of energy needed to
create a perturbation $\zeta(x,t)$ about the mean interface height.
Moreover, as we are assuming a constant surface tension $\gamma$,
\begin{equation}
{\mathcal H}(\{\zeta\})=\gamma\int_{-L/2}^{L/2}dx\left[
\sqrt{1+(\partial_x\zeta)^2}-1\right]\\
\simeq\frac{\gamma}{2}\int_{-L/2}^{L/2}dx \, (\partial_x\zeta)^2,
\end{equation}
provided the perturbation, $\zeta$, is small enough. The distribution
(\ref{Prob_boltz}) leads to a fluctuation spectrum at equilibrium
(for a system with lateral dimension $L\rightarrow\infty$) of the form
\begin{equation}
\langle\zeta_k\zeta_{-k}\rangle=\frac{k_BT}{\gamma k^2}.
\label{spec_zeta}
\end{equation}
Introducing the boundary condition (\ref{n_boundary_cond}) into the equation
for the velocity (\ref{n_velocidad}) and linearizing with respect to $\zeta$,
we obtain the following algebraic equation in Fourier space,
\begin{equation}
\zeta_{{\bf k}\omega}=\frac{\Omega k_D}{i\omega+\Omega k_D\Gamma
k^2}\chi_{{\bf k}\omega},
\end{equation}
where we have neglected the surface diffusion terms. Therefore, integrating
$\langle\zeta_{{\bf k}\omega}\, \zeta_{{\bf k}^\prime\omega^\prime}\rangle$
in $k^\prime$ and $\omega^\prime$ and comparing the ensuing fluctuation
spectrum to Eq.\ (\ref{spec_zeta}) we find
\begin{equation}
I=\frac{2\Gamma k_BT}{\Omega k_D\gamma}=\frac{2c^0_{eq}}{k_D}.
\end{equation}
Finally, in order to calculate $P$ we assume that the fluctuations due to each
relaxation mechanism are independent of one another. In this respect, we take
the chemical potential difference between the interface and
the vapor to be given by~\cite{mullins:1957,mullins:1959}
$\mu=\Omega\frac{\delta{\mathcal H}}{\delta\zeta}$,
where $\delta/\delta\zeta$ denotes functional derivative.
Linearizing the equation for the velocity, and considering only the
contribution due to surface diffusion, we get
\begin{equation}
\partial_t\zeta=\frac{\Omega\nu_s
D_s}{k_BT}\nabla_s^2\mu+\eta_{SD}=\frac{\Omega^2\nu_s D_s}{k_BT}
\nabla_s^2\frac{\delta{\mathcal H}}{\delta\zeta}+\eta_{SD},
\label{velo_SD}
\end{equation}
where $\nabla_s^2$ is the Laplace-Beltrami (surface) Laplacian, and $\eta_{SD}$
is a noise term related to the fluctuations of the surface diffusion current.
To ensure that (\ref{Prob_boltz}) is the equilibrium distribution, this noise term
must satisfy the fluctuation-dissipation theorem~\cite{Landau}, that here reads
\begin{equation}
\langle
\eta_{SD}(x,t)\, \eta_{SD}(x^\prime,t^\prime)\rangle=2
\Omega^2\nu_s D_s\left(-\nabla_s^2\right)\delta(x-x^\prime)\delta(t-t^\prime).
\end{equation}
Therefore, comparing Eq.\ (\ref{velo_SD}) to Eq.\ (\ref{n_velocidad})
we find that $-\Omega\nabla_s\cdot{\bf p}=\eta_{SD}$ and
consequently $P=2D_s\nu_s$.



\section{Linear stability analysis}

Eqs.\ (\ref{n_difusion_eq})-(\ref{C_anode}) provide a full
description of diffusive growth systems including fluctuations. They thus
generalize the classical model of CVD and can also described (through the
appropriate mapping, as seen above) simplified ECD systems. However,
such a stochastic moving boundary problem is very hard to handle for
practical purposes. In this section we will reformulate it into an integro-differential
form that will allow us to derive (perturbatively) an approximate evolution
equation for the interface height fluctuation, $\zeta(x,t)$. In this respect, we
will use a technique based on the Green function theorem which has been
successfully applied to other similar diffusion problems~\cite{Karma,Pierre}.
For brevity, we show here the main results leaving the technical
details for the interested reader in the appendices.

Our point of departure is the integro-differential equation
\begin{equation}
\frac{c({\bf r},t)}{2}=c_a-\int_{-\infty}^{t}dt^\prime\Bigg[
\int_{-\infty}^{\infty}dx^\prime\left( V+
\frac{\partial \zeta^\prime}{\partial t^\prime}\right)c^\prime G-
D \int_{\zeta^\prime}ds^\prime\left( c^\prime\frac{\partial G}{\partial
n^\prime} - G\frac{\partial c^\prime}{\partial
n^\prime}\right)\Bigg]_{z^\prime=\zeta^\prime} -\sigma({\bf r},t),
\label{main_eq}
\end{equation}
where $G$ is the Green function pertinent to the present diffusive problem.
The single Eq.\ \eqref{main_eq} relates the concentration at the boundary
with the surface height, and is shown in Appendix \ref{green_ap} to be
actually equivalent to the full set of equations \eqref{n_difusion_eq}-\eqref{C_anode},
providing essentially the so-called Green representation formula for our
system \cite{haberman}. Unfortunately, Eq.\ \eqref{main_eq} is still highly
nonlinear and has also multiplicative noise (through the noise term
$\sigma$, see App.\ \ref{green_ap}). Notwithstanding, it will allow us to
perform a perturbative study in a simpler way.

First, let us consider solutions of Eq.\ (\ref{main_eq}) that are of the form
$c=c_0+c_1$, where $c_0$ stands for the part associated with the flat
(i.e., $\mathbf{r}$-independent) front solution, and $c_1$ is a small perturbation
of the same order as the height fluctuation $\zeta(x,t)$. Hence, to lowest order
in the latter and its derivatives (see App.\ \ref{zeroth_order}),
\begin{equation}
c_0=\frac{Vc_a+k_Dc^0_{eq}}{V+k_D}.
\label{c0_b}
\end{equation}
One remarkable feature of the Green function representation is that, from the
knowledge of the concentration at the boundary, we can extrapolate the
value of the particle concentration everywhere. Thus, from Eq.\ \eqref{main_eq}
and using (\ref{semiint}) we find
\begin{equation}
c_0(z)=c_a+(c_0-c_a)e^{-zV/D}.
\label{concentracion_z}
\end{equation}
This equation has been theoretically obtained and experimentally
verified by L\'eger {\em et al.}~\cite{Leger2000}.

We now proceed with the next order of the expansion. At this order we already
obtain a proper (albeit linear) evolution equation for the interface, which moreover
contains all the noise terms contributions, that is (see App.\ \ref{first_order}),
\begin{equation}
\partial_t\zeta_k(t)=\omega_k\zeta_{k}(t)+\eta_k(t),
\label{pre_evol_eq}
\end{equation}
where $\omega_k$ is a function (dispersion relation) of wave-vector $k$ whose form
may change with the values of the phenomenological parameters (see below), and
gives the rate at which a periodic perturbation of the flat profile grows (if $\omega_k>0$)
or decays (if $\omega_k<0$) as a function of $k$. Note that, being linear,
Eq.\ \eqref{pre_evol_eq} can be exactly solved.

It turns out that the behavior of $\omega_k$ can be most significantly studied as a function
of the values of the kinetic coefficient $k_D$, as anticipated in Secs. \ref{cvd_sec} and
\ref{ecd_sec}. Specifically, we analyze separately the case in which
surface kinetics is instantaneous (that is, the sticking probability is
high) and all other cases in which the attachment rate is finite.

\subsection{Non-instantaneous surface kinetics ($k_D<\infty$)}

For a finite value of the kinetic coefficient $k_D$, we cannot obtain (even in the zero-noise limit)
the linear dispersion relation $\omega_k$ in a closed analytic form, unless we perform a
large scale ($k\rightarrow 0$) approximation. Thus, we can analyze implicitly the zeros
of the function ${\mathcal T}_{k\omega}$ defined in \eqref{transfer}, which yield the
required form of $\omega_k$ as a function of wave-vector~\cite{tercergrado}. In the
large scale limit we find \cite{MarioTesi}
\begin{equation}
\omega_k=a_2k^2-a_4k^4,
\label{taylor}
\end{equation}
where
\begin{equation}
a_2=\frac{Dk_D}{V}\Delta, \quad a_4=
\frac{D k_D l_D d_0\Delta }{V\bigg[1-\sqrt{\frac{Vd_0}{D}}\bigg]}.
\label{taylor_1}
\end{equation}
where $\Delta=1-d_0/l_D$. The two constants appearing in $\Delta$ are the capillarity length
($d_0\equiv\Gamma\Omega$) and the diffusion length ($l_D\equiv D/V$). If $\Delta<0$,
then $k=0$ is the only zero of $\omega_k$, and since  $a_2<0$ all Fourier modes
$\zeta_k(t)$ of the height fluctuation are stable, since they decay exponentially in time
within linear approximation.

\begin{figure}[!ht]
\includegraphics[width=0.5\textwidth]{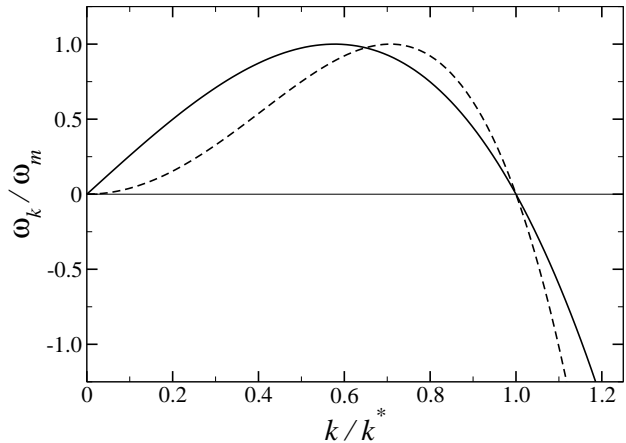}
\caption{Linear dispersion relations given by Eqs.\ (\ref{taylor}) (dashed line) and (\ref{MS_disp})
(solid line), normalized by the growth-rate of the most unstable mode, vs spatial frequency
$k$ normalized by $k^*$. Both axes are in arbitrary units.}
\label{relacion_KS}
\end{figure}

On the contrary, of $\Delta>0$, then $a_2$ and $a_4$ are both positive and there is a band
of unstable modes for all $k \in (0,k^*)$, with
$k^*=\left[\frac{V}{Dd_0}\left(1-\sqrt{\frac{Vd_0}{D}}\right)\right]^{1/2}$.
For these values of the wave-vector, $\zeta_k(t)$ grows exponentially in time within linear
approximation. A maximally unstable mode exists corresponding to the maximum positive
value of $\omega_k$, whose amplitude dominates exponentially all other and leads to the
formation of a periodic pattern.
Under these parameter conditions, the dispersion relation \eqref{taylor} is that of the linear
Kuramoto-Sivashinsky (KS) equation (see Fig.\ \ref{relacion_KS}) \cite{kuramoto}.

Although the above linear dispersion relation contains $O(k^4)$ terms, typical of relaxation
by surface diffusion \cite{mullins:1957,mullins:1959}, these originate as higher order
contributions in which diffusion ($D$), aggregation ($k_D$), and surface tension ($\Gamma$)
become coupled. We can also include proper surface diffusion into the analysis, for which
we simply have to replace $a_4$ by $a_4+B\Omega(V+k_D)/V$, with $B$ as in \eqref{B_Mullins},
which merely shifts $k^*$ closer to zero. In this case, the band of unstable modes shrinks, which
is consistent with the physical smoothing effect of surface-diffusion at short length scales.


\subsection{Instantaneous surface kinetics ($k_D\rightarrow\infty$)}

If the sticking probability is essentially one, as seen above $k_D\rightarrow\infty$.
This fast attachment condition occurs in many irreversible growth processes \cite{meakin}.
Following a similar procedure (and long wavelength approximation) as the one that led
us to the KS dispersion relation in the previous section, we now get
\begin{equation}
\omega_k=D\left(\frac{\Gamma^2\Omega^2}{2}- \frac{B\Omega}{D}\right)k^4
-\frac{3\Gamma\Omega Vk^2}{2}\nonumber+|k|(V-\Gamma\Omega Dk^2)
\bigg[1-\frac{\Gamma\Omega V}{D}+
\left(\frac{\Gamma^2\Omega^2}{4}-
\frac{B\Omega}{D}\right)k^2 \bigg]^{1/2}.
\label{relacion_disp_inf}
\end{equation}
This expression has several interesting limits. For instance, if we
neglect surface tension and surface diffusion terms (that is, for $\Gamma=B=0$),
then $\omega_k=V|k|$, the well-known dispersion relation of the Diffusion Limited
Aggregation (DLA) model \cite{meakin}. In this case, every spatial length scale is unstable,
the shortest ones (large $k$ values) growing faster than the larger ones. Thus, in such
a case the aggregate consists of wide branches plenty of small tips. Moreover, there is
actually no characteristic length scale in the system, hence the aggregate has scale invariance (that is, it is self-similar).

If we only neglect the surface diffusion term and, since $\Gamma\Omega\equiv d_0$
(the capillarity length) is typically in the range $10^{-7}-10^{-6}$ cm, and $D/V\equiv l_D$
(the diffusion length) is close to $10^{-1}-10^{-2}$ cm, we can write
\begin{equation}
\omega_k\simeq V|k|(1-d_0l_Dk^2),
\label{MS_disp}
\end{equation}
which is the celebrated Mullins-Sekerka (MS) dispersion relation~\cite{Mullins_Sekerka1,Mullins_Sekerka2}
(see Fig.\ \ref{relacion_KS}), ubiquitous in growth systems in which diffusive instabilities
(induced by shadowing of large branches over smaller surface features) compete with relaxation
by surface tension. This dispersion relation has been experimentally verified in several ECD
systems~\cite{Bruyn,Kahanda,Tesis_Juanma}, and has actually been theoretically proposed
before for ECD by Barkey {\em et al.}~\cite{Barkey} (although under non-galvanostatic conditions).

However, in many diffusive growth systems both surface tension and surface diffusion are non-negligible;
considering again the physical hypothesis $d_0\ll l_D$ and a long-wavelength approximation, we get
\begin{equation}
\omega_k=V|k|[1-(d_0l_D+B\Omega/2D)k^2]-B\Omega k^4.
\label{MS+SD}
\end{equation}
Nevertheless, there are $e.g.$ some CVD conditions \cite{Ojeda,Ojeda:2003}, for which the vapor pressure
in the dilute phase is so low that relaxation by evaporation/condensation is negligible in practice.
In such a case, the dispersion relation is provided by \eqref{MS+SD} with an effective zero value for $d_0$.

Finally, there may be physical situations in which
quite analogously to the KS case seen above, the last dispersion relations [Eqs.\ \eqref{MS_disp}
through \eqref{MS+SD}] show the competition between mechanisms which tend to destabilize the
interface and other which tend to stabilize it. From this competition, a characteristic length-scale arises,
$\lambda_m$, which grows exponentially faster than the others ($\lambda_m=2\pi/k_m$, with $k_m$
being the value for which $\omega_k$ is a positive maximum).

In summary, we see that, in reducing the efficiency of attachment from complete ($k_D\to\infty$) to finite
($k_D<\infty$), the symmetry of the dispersion relation  changes so that non-local terms like odd powers
of $|k|$ are replaced by local (linear) interactions. For instance, $-k^2 \zeta_k(t)$ is the Fourier transform
of the local term $\partial_x^2 \zeta(x,t)$, while $|k| \zeta_k(t)$ cannot be written as (the transform of)
any local differential operator acting on $\zeta(x,t)$. This result can be understood heuristically: if the
sticking probability is small, then particles arriving at the interface do not stick to it the first time they
reach it, but they can explore other regions of the aggregate. This attenuates the non-local shadowing
effect mentioned above, so that growth becomes only due to the local geometry of the surface.

Although the $k_D\rightarrow\infty$ limit is a mathematical idealization, for practical purposes we can
determine under which conditions it is physically attained. Thus, if we introduce Eq.\ (\ref{concentracion_z})
describing the concentration field for a flat interface into the boundary condition (\ref{n_boundary_cond}),
the latter takes the form
\begin{equation}
\frac{c-c^0_{eq}}{D/k_D}=\frac{c-c_0}{D/V}.
\label{criterio_sticking}
\end{equation}
The term $D/V$ is the diffusion length; hence, analogously, we can define
$D/k_D$ as a {\em sticking length}. Physically, this length can be seen as the
typical distance traveled by a particle between its first arrival at the
interface and its final sticking site. We can neglect this length scale
if the sticking probability is close to unity. On the contrary, if
$k_D\rightarrow 0$, the distance that the particle can explore before attaching is infinite.
Therefore, taking Eq.\ (\ref{criterio_sticking}) into account, we can say that the $k_D\rightarrow\infty$
limit describes accurately the problem when $k_D\gg V$, and in such a case the diffusion length and
the capillarity length determine the characteristic length-scale of the system.

\section{Nonlinear evolution equation}
\label{nee}

In this section we proceed one step further with our perturbative approach by including the lowest
order nonlinear contributions to Eq.\ (\ref{pre_evol_eq}). If we evaluate the Green function of the
problem at the boundary we find
\begin{equation}
G({\bf r}-{\bf r}^\prime,\tau)=\frac{\Theta(\tau)}{4\pi D\tau}
\exp\left[-\frac{(x-x^\prime)^2}{4D\tau}-
\frac{\big(\zeta-\zeta^\prime+V\tau\big)^2}{4D\tau}\right],
\label{nl_green}
\end{equation}
where $\tau=t-t^\prime$. Expanding the last term in the argument of the exponential as a series in
$\zeta$ we get $(\zeta-\zeta^\prime)^2+2V(\zeta-\zeta^\prime)\tau+V^2\tau^2.$
The second and third terms were already taken into account before in the linear analysis, so the
only nonlinear contribution in Eq.\ (\ref{nl_green}) is related to the first term, $(\zeta-\zeta^\prime)^2$,
that introduces a correcting factor
$\exp [-(\zeta-\zeta^\prime)^2/(4D\tau)]$
which is only significant when $\zeta^\prime\simeq\zeta$,
hence we can replace $\zeta^\prime-\zeta$ by the lowest term in its Taylor
expansion, to get
\begin{equation}
\exp\left(-\frac{(\zeta-\zeta^\prime)^2}{4D\tau}\right)\simeq 1-
\frac{(\partial_x\zeta)^2(x^\prime-x)^2}{4D\tau} .
\end{equation}
By incorporating this contribution into the formulae of appendices \ref{green_ap}
through \ref{first_order}, the consequence can be readily seen to be the addition of a mere term
equal to the (Fourier transform of)
$V(\partial_x\zeta)^2/2$ to the right hand side of Eq.\ \eqref{pre_evol_eq}, resulting into an
evolution equation with the form
\begin{equation}
\partial_t\zeta_k(t)=\omega_k\zeta_{k}(t)+ \frac{V}{2}\mathcal{N}[\zeta]_k+\eta_k(t),
\label{pre_KPZ}
\end{equation}
where $\mathcal{N}[\zeta]_k$ is the Fourier transform of $\mathcal{N}[\zeta]=(\partial_x\zeta)^2$.
Note that the nonlinear term obtained is precisely the characteristic KPZ nonlinearity as in
Eq.\ (\ref{kpzeq}), that appears here with a coefficient equal to half the average growth velocity,
agreeing with standard mesoscopic arguments \cite{KPZ}. Moreover, as is well known
\cite{Barabasi}, this is the most relevant nonlinear term (asymptotically) that can be obtained
for a non-conservative growth equation such as Eq.\ (\ref{main_eq}), hence any other nonlinear
term will not change the long time, long scale behavior of the system. As we have shown above,
for small sticking $\omega_k$ is given by Eq. \eqref{taylor} and for large sticking is given by
Eq. \eqref{MS+SD}.
In all cases the noise correlations involve constant terms, as well as terms that are proportional to
successively higher powers of $k$. By retaining only the lowest order contributions in a long-wavelength
and quasistatic  approximation, from Eq.\ (\ref{betabeta}) we obtain
\begin{equation}
\label{eta_KS}
\langle\eta_{k\omega}\eta_{k^\prime\omega^\prime}\rangle=
(\Pi_0+\Pi_2k^2)\delta(k+k^\prime)\delta(\omega+\omega^\prime),
\end{equation}
where
\begin{equation}
\Pi_0 = \left\{ \begin{array}{lc}
V c_0 (1 + \frac{2V}{k_D}), & k_D < \infty \\
V c_{eq}^0, & k_D \to \infty
\end{array} \right. , \label{pi0}
\end{equation}
and
\begin{equation}
\Pi_2 = \left\{ \begin{array}{lc}
2 D^2 c_0 (\frac{1}{V}+\frac{2}{k_D}) + 2D_s \nu_s (1+\frac{V}{k_D})^2, & k_D < \infty \\
2 D^2 c_{eq}^0/V + 2D_s \nu_s, & k_D \to \infty
\end{array} \right. \label{pi2}
\end{equation}
where Eq.\ \eqref{c0_b} for $c_0$ is to be used in the case of finite kinetics. In general, parameter $\Pi_0$
provides the strength of non-conserved noise, while $\Pi_2$ measures the contribution of conserved
noise \cite{GarciaOjalvo} to the interface fluctuations. The presence of conserved noise can naturally
modify some short length and time scales of the system but, in the presence of non-conserved noise,
it is known to be irrelevant to the large scale behavior. Thus, $\Pi_2$ will be neglected in the numerical
study performed below.

For the case of finite kinetic coefficient, the evolution equation \eqref{pre_KPZ} is the stochastic
generalization of the KS equation \cite{cuerno:1995,cuerno:1995b,Ueno:Kuramoto}.
In the case of fast attachment $k_D\to\infty$, the resulting interface equation
combines the linear dispersion relation of MS with the KPZ nonlinearity. In this sense it employs two
``ingredients'' that seem ubiquitous in growth systems, so we find it remarkable the fact that (to the
best of our knowledge) its detailed dynamics has not been reported so far.
The purpose of the next section is to report a numerical study of
this equation
in order
to clarify the similarities and differences to its finite attachment counterpart
.

\subsection{Numerical results}

\subsubsection{The pseudo-spectral method}
As noted above, although the shape of the nonlinear
Eq.\ (\ref{pre_KPZ})
is common to both sticking limits, their  dispersion
relations make them very different physically.
Thus, while the linear terms of the KS equation are local in space,
linear terms corresponding to MS dispersion relation
cannot be written in terms of local
spatial derivatives.
Therefore, we cannot perform
a standard finite difference discretization in order to implement its
numerical integration~\cite{GarciaOjalvo}.
Rather, in order to integrate numerically
Eq.\ (\ref{pre_KPZ})
we
will resort to the so-called pseudo-spectral methods that make use both of real and Fourier space
representations. Such techniques have been successfully used $e.g.$ in many instances of the
Physics of Fluids~\cite{Canuto}, and are being used more recently in the study of stochastic partial
differential equations~\cite{Giacometti,Toral,giada:2002a,giada:2002b,gallego}.


As we are interested in the qualitative scaling properties of Eq.\ (\ref{pre_KPZ}),
rather than, say, in a quantitative comparison to a specific physical system, we introduce
positive constants $\nu$, $K$, $B$, and $\lambda$, that allow us to write the equations
in the more general form for each limit:
\begin{eqnarray}
\partial_t\zeta_k(t)&=&(\nu k^2-Kk^4)\zeta_k+ \frac{\lambda}{2}\mathcal{N}[\zeta]_k+
\eta_k(t),\label{KS_p} \\
\partial_t\zeta_k(t)&=&(\nu|k|-K|k|^3)\zeta_k+\frac{\lambda}{2}\mathcal{N}[\zeta]_k+
\eta_k(t),\label{MS_p}
\end{eqnarray}
where, in order to stress the similarities between the two interface equations,
we have neglected in \eqref{MS_p} the $O(k^4)$ terms, given that their  stabilizing
role is already played by the $O(|k|^3)$ term.
In what follows, we will only refer to the new parameters, which will later
be estimated in the next section when we analyze different experimental
conditions.

In order to integrate efficiently Eq. \eqref{pre_KPZ} we use a pseudo-spectral scheme.
This numerical
method is detailed in \cite{gallego} and employs an auxiliary change of variable that
allows to estimate the
updated value of $\zeta_k$
as
\begin{equation}
\zeta_k(t+\Delta t)=e^{\omega_k \Delta t}\left( \zeta_k(t)+\frac{\lambda}{2}\Delta t\,
	{\mathcal N}[\zeta(t)]_k\right)+r_k(t),
\label{pseudo_scheme}
\end{equation}
where
the noise term $r_k(t)$ 
is conveniently expressed as
\begin{equation}
r_k(t)=\sqrt{\left(\Pi_0+\Pi_2 k^2\right)
	\frac{e^{2\omega_k\Delta t}-1}{2\omega_k\Delta x}}\, v_k(t),
\end{equation}
with $v_k(t)$ being the Fourier transform of a set of Gaussian random numbers
with zero mean and unit variance~\cite{Toral}.
Regarding the explicit calculation of the nonlinear term in Eq.\ \eqref{pseudo_scheme},
we perform the inverse Fourier transform of $-ik\zeta_k$ and take the square of it in
real space, so that we explicitly avoid nonlinear discretization issues \cite{Giacometti}.
However, in this procedure aliasing issues arise \cite{Canuto}, that we avoid by
extending the number of Fourier modes involved in the integration, and using zero-padding,
see details in \cite{Canuto,Giacometti}.

\subsubsection{Numerical integration}

\label{resul_num}

As mentioned, the noisy KS equation, Eq.\ (\ref{KS_p}), has been extensively studied
in the literature, so we will concentrate in this section on the novel Eq.\ (\ref{MS_p}).
Its behavior will help us to understand the evolution of the interface in the
$k_D\rightarrow\infty$ cases.

\begin{figure}[!ht]
\includegraphics[width=0.5\textwidth]{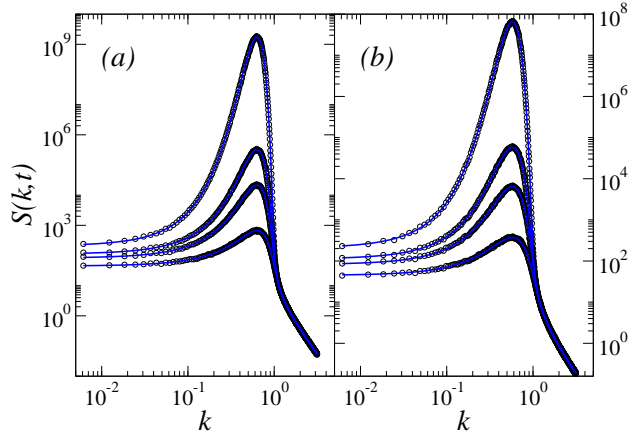}
\caption{(color online) Power spectral density, $S(k,t)$, from numerical simulations for
a $L=1024$ system with $\nu=K=1$, $\Pi_0=10^{-2}$, $\lambda=0$ ($i.e.$, linearized equations),
averaged over $10^3$ realizations, at times $t=4$, $8$, $10$ and $20$ for:
(a) Eq.\ (\ref{KS_p}) (circles) and (b) Eq.\ (\ref{MS_p}) (circles). Blue solid lines represent the
exact solution, Eq.\ (\ref{esp_lineal}), for each case. All axes are in arbitrary units.}
\label{comp_MS_KS}
\end{figure}

\begin{figure}[!ht]
\includegraphics[width=0.5\textwidth]{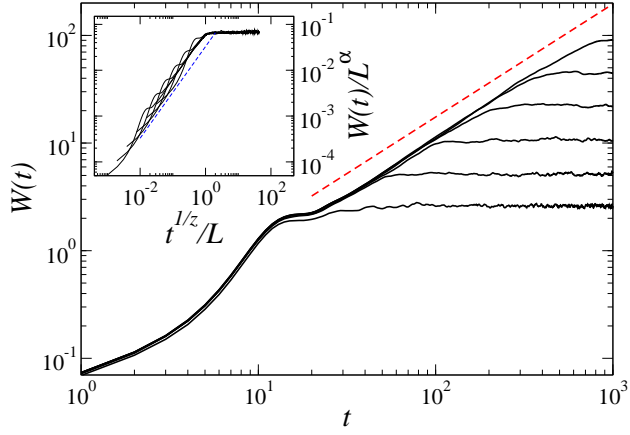}
\caption{(color online) Global width vs time for a system with $\nu=K=\lambda=1,
\Pi_0=10^{-2}$ obtained numerically for Eq.\ (\ref{MS_p}), for increasing system
sizes, $L=32$, $64$, $128$, $256$, $512$ and $1024$, bottom to top, averaged
over $10^3$ realizations. The red dashed line is a guide to the eye with slope $1.05$
suggesting the asymptotic value of $\beta$. Inset: Collapse of $W(t)$ using
$\alpha=1.00$, $\beta=1.05$, and $z=0.95$. The blue line has slope $1.00$, showing
the consistency of our estimate for $\alpha$. Axes in the main panel and in the inset
are all in arbitrary units.}
\label{MS_anchura}
\end{figure}

\begin{figure}[!ht]
\includegraphics[width=0.5\textwidth]{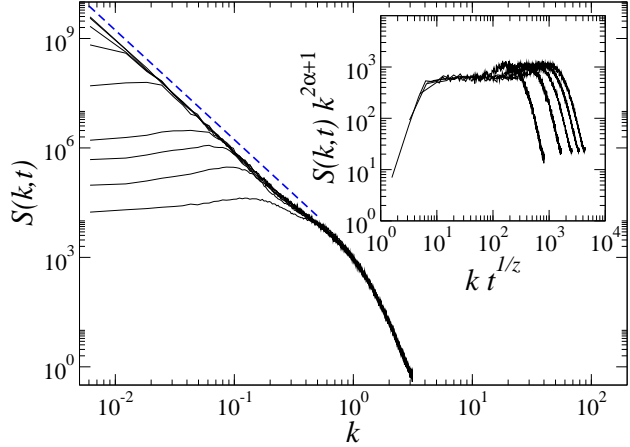}
\caption{(color online) Power spectrum vs spatial frequency $k$ with the same
parameters as in Fig.\ \ref{MS_anchura} and $L=1024$. Different curves
stand for different times (the one at the bottom is for the earliest time).
The dashed line is a guide to the eye with slope $-3$, compatible with $\alpha=1$.
Inset: Collapse of $S(k,t)$ using $\alpha=1.00$, $\beta=1.05$, and $z=0.95$.
Axes in the main panel and in the inset are all in arbitrary units.
}
\label{MS_psd}
\end{figure}

The linear regime is very similar for both equations, as one might naively expect.
In fact, we find that both systems feature similar power spectral densities (or surface
structure factors), $S(k,t) = \langle \zeta_k(t) \zeta_{-k}(t) \rangle$, see Fig.\ \ref{comp_MS_KS}.
This can be easily understood by inspection of the analytic result for $S(k,t)$ obtained
from the exact solution of Eq.\ (\ref{pre_evol_eq}),
\begin{equation}
S(k,t)=\Pi_0\frac{e^{2\omega_k t}-1}{2\omega_k},
\label{esp_lineal}
\end{equation}
where the corresponding dispersion relations are displayed in Fig.\ \ref{relacion_KS}.
By simple inspection of Fig.\ (\ref{comp_MS_KS}) one is tempted to say that
both Eqs.\ (\ref{KS_p}) and (\ref{MS_p}) have a similar behavior so that, {\em a priori},
KPZ scaling might be expected in the asymptotic regime also for the latter. However,
for this fast kinetics equation, we show in Fig.\ \ref{MS_anchura} that the growth exponent
$\beta$ characterizing the power-law growth of the surface roughness or global width
$W(t) \sim t^{\beta}$, where $W^2(t) = (1/L^2) \sum_k S(k,t)$ \cite{Barabasi,cuerno:2004},
is much larger at long times than the expected KPZ value $\beta_{KPZ}=1/3$. A rationale
for such a long-time behavior of Eq.\ (\ref{MS_p}) can be already provided by simple
dimensional analysis. Thus, under the scale transformation $t\to b^zt$, $k\to b^{-1}k$,
$\zeta_k\to b^{\alpha+1}\zeta_k$, Eq.\ \eqref{MS_p} becomes
\begin{equation}
\partial_t\zeta_k=b^{z-1}\nu|k|\zeta_k-b^{z-3}K|k|^3\zeta_k+
b^{\alpha+z-2}\frac{\lambda}{2}\mathcal{N}[\zeta]_k+b^{z/2-\alpha-1/2}\eta_k.
\end{equation}
If we introduce the exact one dimensional KPZ exponents ($\alpha=1/2$, $z=3/2$),
it is easy to see that in the hydrodynamic limit (that is, when $b\rightarrow\infty$) the
most relevant term in the equation is {\em not} the KPZ term but, rather, the lowest
order linear term $|k|\zeta_k$. Preliminary dynamic renormalization group
calculations~\cite{nicoli:2008} seem to provide the same result. The present scaling
argument provides moreover the exponent values $\alpha=\beta=z=1$ at the stationary
state. These values are compatible with those obtained from numerical simulations of
Eq.\ (\ref{MS_p}), as displayed in Figs.\ \ref{MS_anchura} and \ref{MS_psd}. The
numerical values we obtain for the exponents are $\alpha=1.00\pm0.05$, $\beta=1.05\pm0.05$
and $z=0.95\pm0.05$, thus they are in good agreement with the ones predicted by
dimensional analysis. In order to check the consistency of our numerical estimates
\cite{lopez:1997a,lopez:1997b}, in the inset of Fig.\ \ref{MS_psd} we show the collapse
of the power spectrum density using these exponent values. Collapses are satisfactory,
including the behavior of the scaling function for the width,
indicated by a solid line in the inset of Fig.\ \ref{MS_anchura}. The discrepancies in
the collapsed curves for large $kt^{1/z}$ values is due to the existence of a short
scale scaling different from the asymptotic one.

\begin{figure}[!ht]
\includegraphics[width=0.5\textwidth]{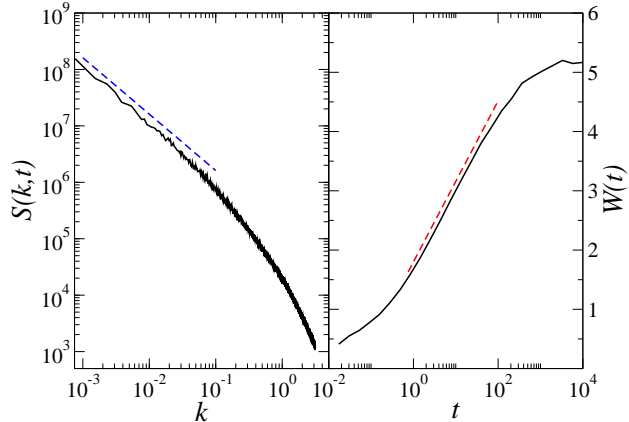}
\caption{(color online) Left: Power spectral density vs wave vector for Eq.\ (\ref{MS_p})
with $\nu=-1$, $K=1$, $\lambda=4$, $\Pi_0=10$, $L=8192$, and averages over
100 noise realizations. The dashed line is a guide to the eye with slope
$-1$ [that is, $\alpha=0$ (log)].
Right: Global roughness vs time for the same system as in the left panel. Note the
semilog representation. The dashed line is a guide to the eye representing
$W(t) \sim \log t$ before saturation. All axes are in arbitrary units.}
\label{fig_nueva}
\end{figure}

From our numerical results, we conclude that the non-linear regime as described
by Eq.\ \eqref{MS_p}, corresponding to instantaneous kinetics, is very different
from that for slow kinetics, as represented by the noisy KS equation. Thus, in both
cases the KPZ nonlinearity is able to stabilize the system and induce power law
growth of the surface roughness, associated with kinetic roughening properties.
However, the universality class of Eq.\ \eqref{MS_p} is {\em not} that of the KPZ
equation but, rather, it is a new class completely determined by the $|k|$ term in
the linear dispersion relation. Note moreover that the new exponents associated
with the asymptotic regime for this equation fulfill {\em accidentally} the Galilean
scaling relation $\alpha+ z =2$ \cite{Barabasi,KPZ}. We describe this property as
accidental due to the fact that Eq.\ \eqref{MS_p} is {\em not} Galilean invariant.
The easiest way to confirm this is to check for the scaling behavior of a {\em stable}
version of Eq.\ \eqref{MS_p} in which we take {\em negative} $\nu$ values for which
all modes are linearly stable. Fig.\ \ref{fig_nueva} shows on the left panel the power
spectral density in this case,
that behaves as $S(k)\sim 1/k$ for long distances. Moreover, on the right panel of
this figure we plot the time evolution of the global roughness for the same system,
that behaves as $W(t) \sim \log t$ for long times before saturation. From these
data we conclude the exponent values are $z=1$, and $\alpha=0$ (log), which do
{\em not} satisfy the scaling relation implied by Galilean invariance.
Such a stabilized version of Eq.\ \eqref{MS_p} has been studied in the
context of  diffusion-limited erosion \cite{krug:1991}.


\section{Comparison with experiments and discrete models}

In this section, we focus on the applications of the model equations
to understand and explain some experimental results, qualitatively and
quantitatively and, besides, compare our results from continuum theory to
relevant discrete models of diffusion limited growth.


\subsection{Experiments}

From the experimental point of view, it is very difficult to determine if the dispersion
relation characterizing a specific physical system is the MS or, rather, the KS one,
because they are both very similar except very near $k=0$
(see Fig.\ \ref{relacion_KS}). In principle, such a distinction would be very informative,
since it could provide a method to assess whether the dynamics is diffusion limited
($k_D\to\infty$), or else reaction limited ($k_D \ll V$). Several previous theoretical
works in this field~\cite{Chen,Elezgaray} predict, qualitatively, a KS dispersion
relation, while other~\cite{Barkey}, predict a dispersion relation similar to the one
of MS. All those studies are not necessarily incompatible with one another because
they are considering different experimental conditions. Moreover, to our knowledge,
there are only a few experimental reports in which the dispersion relation is
measured, and in most cases the authors presume that it can be accounted for
by MS~\cite{Bruyn,Kahanda,Tesis_Juanma}. Notwithstanding, within experimental
uncertainties that are specially severe at small wave-vectors, they could all equally
have been described by the KS dispersion relation.

Another way to distinguish which is the correct effective interface equation that
describes a given system could be the value of the characteristic length associated
with the most unstable mode that can be measured (which in fact would be
essentially the characteristic length-scale that could observed macroscopically for the system).
Unfortunately, from Eq.\ (\ref{taylor_1})
one has that, for $d_0\ll l_D$, then $k_m=(2l_Dd_0)^{-1/2}$ for the KS and
$k_m=(3l_Dd_0)^{-1/2}$ for the MS dispersion relation.
Hence, except for a constant numerical value of order unity, both cases provide a
characteristic length scale that depends equally on physical parameters, while the
significant parameter, $k_D$, only modifies $\omega_m$, namely, the characteristic
time at which the instability appears (which is about $\omega_m^{-1}$). Thus, in
order to clarify the nature of the growth regime (diffusion or reaction limited), one
should rather study the long time behavior of the interface.

Despite these difficulties, we can sill try to interpret some experimental
results reported in the literature. L\`eger {\em et al.}~\cite{Leger,Leger2,Leger2000}
have presented several exhaustive works dealing with ECD of Cu under galvanostatic
conditions. In addition to properties related to the aggregate, they also provide
detailed information about the cation concentration. Their main result in this respect
is that such concentration obeys experimentally Eq.\ (\ref{concentracion_z}) as we
anticipated above. It can also be seen that the product aggregates are branched
and that the topmost site at every position $x$ defines a rough front growing with a
constant velocity.
In Fig.\ 8 of Ref.\ \cite{Leger2000} the authors plot the branch width against the diffusion length.
From that figure, it is clear that $\lambda_m$ and $l_D$ are not linearly related. Moreover,
they seem to better agree with the present prediction from either of our effective interface equations,
$\lambda_m\propto l_D^{1/2}$, than with the linear behavior argued for in \cite{Leger2000}.

Another important experimental feature is the relation between $\lambda_m$ and $C_a$.
From Eq.\ (\ref{Omega_c_a}) we know that $\Omega=(1-t_c)/C_a$,
then $d_0=\Omega\Gamma=\Omega\gamma/RT$, so that the characteristic length
scale, $\lambda_m$, will be proportional to $C_a^{-1/2}$.
Thus, one would expect
the branches to be narrower as we increase the initial concentration,
consistent with the patterns obtained by L\`eger {\em et al.}

We will try also to interpret the ECD experiments of Schilardi {\em et al.}~\cite{Schilardi}.
They have measured the interface global width
considering that the topmost heights of the branches provide a well defined front. Thus,
the time evolution of the global width, or roughness, presents three different well defined
regimes: A short initial transient, which cannot be accurately characterized by any power-law
due to the lack of measured points, is followed by an unstable transient. We consider the
system unstable in the sense that the average interface velocity is {\em not} a constant
but, rather, grows with time. Finally, the system reaches a regime characterized by
exponent values that are compatible with those of KPZ the equation, while the aggregate
grows at a constant velocity. These regimes again resemble qualitatively the behavior
expected for the noisy KS equation (\ref{KS_p}).

Moreover, we can check whether the order of magnitude of the experimental parameters
is compatible with our predictions. First, we estimate $k_m$ from
mean width of the branches within the unstable regime. This width is about $0.05$ mm~\cite{Schilardi},
hence $k_m\simeq 1.3\times 10^{3}$ cm$^{-1}$. Besides this, the mean aggregate velocity at
long times is $V\simeq 2\times 10^{-4}$ cm\,s$^{-1}$, and the diffusion coefficient is
$D\simeq 10^{-5}$ cm$^2$\,s$^{-1}$, so that the diffusion length is about $l_D=D/V\simeq 0.05$ cm.
These magnitudes allow us to calculate the capillarity length
$d_0=1/2l_Dk_m^2\simeq 5\times 10^{-6}$ cm (which is of the same order as we considered in
our approximations above, and much smaller than the diffusion length). Furthermore, the
instability appears at times of order $1/\omega_m$. In the experiment this time is about $6$ min.
Thus, $\omega_m\simeq 3\times 10^{-3}$ s$^{-1}$. Finally, as $\omega_m=k_Dl_Dk_m^2/2$,
then $k_D\simeq 6\times 10^{-8}$ cm\,s$^{-1}\ll V$, which provides a consistency criterion for
the validity of our approximations and of our predictions.

As a final example, let us perform some comparison with the mentioned
experiments of Pastor and Rubio~\cite{Pastor,Tesis_Juanma}. At short times,
they obtain compact aggregates with exponent values~\cite{Tesis_Juanma}
$\alpha=1.3\pm 0.2$, $\alpha_{loc}=0.9\pm 0.1$, $z=3.2\pm 0.3$, and
$\beta=0.4\pm 0.08$. This means that the interface is superrough ($\alpha>1$).
After this superrough regime, the aggregate becomes unstable and the
dispersion relation has the MS form. The fact that the aggregates are compact (and not ramified)
seems to show that surface diffusion is an important growth mechanism in these experiments
(which is also consistent with the small value of the velocity, $V\simeq 4$~$\mu$m/min).
Consequently, we can use Eq.\ (\ref{MS_p}) with an additional surface diffusion contribution
[$-Bk^4\zeta_k(t)$] to understand the behavior of the experiments. We choose the parameters
$\nu=0.25$, $K=\Pi_0=1$ and $\lambda=\Pi_2=0$ (because the velocity is small and we are
only interested in the short time regime), for several values of $B$  between $0$ and $1$ in
order to determine the influence of surface diffusion in the growth exponents. Other parameters
only change the characteristic length and time scales of the experiment. The exponents thus
obtained numerically are (with $B=0.75$) $\beta\simeq 0.39\pm 0.02$, and
$\alpha\simeq 1.3\pm 0.1$,
which are (within error bars) equal to the experimental ones. As we have pointed out above,
this superrough regime is followed by an unstable transient characterized by the MS
dispersion relation, as has been also observed in other ECD experiments by de
Bruyn~\cite{Bruyn}, and Kahanda {\em et al.}~\cite{Kahanda}.

More recently, additional ECD experiments have been reported under galvanostatic
conditions. $E.g.$ in Ref.\ \cite{lafouresse} three growth regimes can be distinguished:
a first one at short times, in which a Mullins-Sekerka like instability is reported, is followed
by a regime in which anomalous scaling (namely, the roughness exponents measured
from the global and local surface widths differ, $\alpha\neq \alpha_\textrm{loc}$)
\cite{lopez:1997a,lopez:1997b,cuerno:2004} takes place, and finally at long times
ordinary Family-Vicsek scaling \cite{Barabasi} is recovered. Similar transitions to and
from anomalous scaling behavior have been also reported in Ref.\ \cite{osafo}.
Our present theory does not predict the anomalous scaling regimes reported by
these authors. This may be due to the small slope condition employed in the
derivation of equations (\ref{MS_p}) and (\ref{KS_p}). However, we want to emphasize
two points in this respect:
{\em (i)} In Ref.\ \cite{osafo} the authors report a transition from rough interface behavior
to mound formation. These mounds can be obtained by numerical integration of equation
(\ref{MS_p}) in 2+1 dimensions \cite{castro:2008}.
{\em (ii)} As we will show in the next section, the theory is in good agreement
with a discrete model of growth in which anomalous scaling is clearly
reproduced.
Hopefully, a numerical integration of the full moving boundary problem (\ref{main_eq})
would capture the anomalous scaling regime, and this will be the subject of further work.

\subsection{Discrete Models}

As mentioned above, $k_D$ is related with the sticking probability through Eq.\ (\ref{k_D_sticking}).
This probability acts as a noise reduction parameter~\cite{noise_reduction} in discrete growth models,
as was shown $e.g.$ in \cite{mbdla1,mbdla2,mbdla3} for the Multiparticle Biased Diffusion Limited
Aggregation (MBDLA) model, used to study ECD growth. In particular, MBDLA has been seen to
describe quantitatively the morphologies obtained in \cite{Schilardi}. In MBDLA, by reducing the
sticking probability the asymptotic KPZ scaling is indeed
more readily achieved, reducing the importance of pre-asymptotic unstable
transients, as illustrated by Fig.\ 8 of \cite{mbdla3}. Hence, noise reduction is not a mere computational
tool for discrete models but, rather, it can be intimately connected with the surface kinetics via
Eqs.\ (\ref{k_D_sticking}) and (\ref{n_boundary_cond}).

\begin{figure}[!ht]
\includegraphics[width=0.5\textwidth]{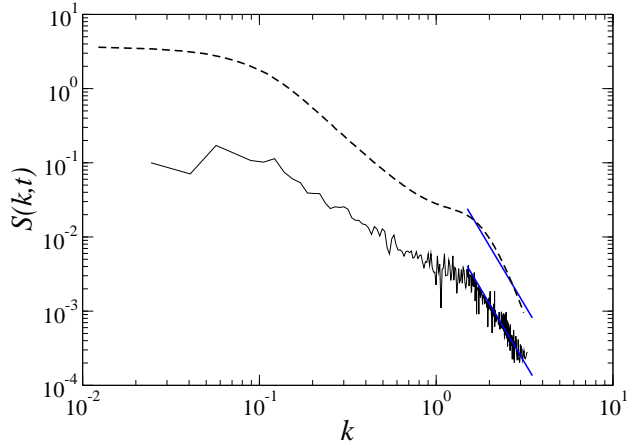}
\caption{(color online) Power spectrum obtained from MBDLA simulations (solid line) with
$r=0.5$, $p=0.5$, $s=1$, $c=1$ (taken from Fig.\ 15 in \cite{mbdla3}, by permission), and
Eq.\ (\ref{KS_p}) with $\nu=1$, $K=1/4$, $\lambda=40$, $\Pi_0=10^{-2}$, $L=512$ (dashed line),
averaged over $10^5$ realizations. For the sake of clarity, the latter has been vertically offset.
Straight blue lines are guides to the eye having slope $-4$. Axes are in arbitrary units.}
\label{KS_vs_bollo}
\end{figure}

MBDLA with surface diffusion also predicts the existence of a characteristic branch width.
This is shown in Fig.\ \ref{KS_vs_bollo} in which the power spectral density is plotted and
compared with the one obtained from the noisy KS equation, proving the equivalence
between both descriptions of ECD.
These results are reinforced by the fact that, as shown in Ref.\ \cite{mbdla3},
the cation concentration obeys Eq.\ (\ref{concentracion_z}), and the branch width dependence on
the cation concentration is consistent with the relation $\lambda_m\propto C_a^{-1/2}$. Moreover,
in simulations of MBDLA, an unstable transient was found before the KPZ scaling regime, being
characterized by intrinsic anomalous scaling, as recently observed in the experimental works by
Huo and Schwarzacher \cite{SchwarLett1,SchwarLett2}. As mentioned, probably the absence of
such an anomalous scaling transient in our continuum model is
related to the small slope approximation, and we expect to retrieve it from a numerical integration
of the full moving boundary problem.

\section{Conclusions}

In this paper we have provided a derivation of stochastic interface equations from the basic
constitutive laws that apply to growth system in which aggregating units are subject to diffusive
transport, and attach only after (possibly) finite reaction kinetics. Our derivation seems to be
new for this class of systems, and allows to relate the coefficients in the effective interface
equation with physical parameters, like the sticking parameter, physical surface tension and
size of aggregating units, etc.
We have seen that the shape of the equation describing the time evolution of the aggregate
interface changes as a function of the sticking probability. For very high interface kinetics
non-local shadowing effects occur, while for finite kinetics non-local shadowing yields to
morphological instabilities of a local nature. Thus, qualitatively the behavior of the system for
generic parameter conditions roughly consists of an initial transient associated with morphological
instabilities in which typical length scales are selected (thus breaking scale invariance), that is
followed by a late time regime in which the interface displays kinetic roughening. However, the
universality class of the latter differs, being of the KPZ class only for finite attachment kinetics,
while it becomes of a new, different, non-KPZ type for infinitely fast attachment,
as predicted by the new interface equation we obtain in the latter condition. While in previous
reports \cite{ourprl,cuerno:2004,cuerno:2007} we have interpreted the long unstable transients
as a potential cause for the experimental difficulty in observing KPZ scaling, our present results
go one step further in the sense that for fast attachment conditions we do not even expect
KPZ universality in the asymptotic state, due to the irrelevance of the KPZ nonlinearity as
compared with the $|k| \zeta_k(t)$ term in Eq.\ \eqref{MS_p}.

Comparison of our continuum model with experiments and discrete models seem to support
the above conclusions. Nevertheless, our results are in principle constrained by a small-slope
approximation. In view (specially under the fast kinetics conditions) of the large roughness
exponent values that characterize the long time interfaces as described by our effective interface
equations, it is natural to question whether the same scenario holds for the full (stochastic)
moving boundary problem. Moreover, our small slope equations do not account for anomalous
scaling, which is otherwise seen in experiments and in the MBDLA model, so that integration
of the complete system \eqref{n_difusion_eq}-\eqref{C_anode} seems indeed in order.
Technically, systems of this type pose severe difficulties even to numerical simulations
(mostly related to front tracking in the face of overhang formation). Thus, one needs to
rephrase the original continuum description \eqref{n_difusion_eq}-\eqref{C_anode} into
an equivalent formulation that is more amenable to efficient numerical simulation, such as
$e.g.$ a phase-field model \cite{gonzalez-cinca:2004}. We are currently pursuing such
type of approach, and expect to report on it soon.

\acknowledgments

This work has been partially supported by UC3M/CAM (Spain) Grant No.\ UC3M-FI-05-007,
CAM (Spain) Grant No.\ S-0505/ESP-0158, and by MEC (Spain), through
Grants No.\ FIS2006-12253-C06-01, No.\ FIS2006-12253-C06-06, and the FPU program (M.\ N.).

\appendix

\section{The Green function technique}
\label{green_ap}

\begin{figure}[!ht]
\includegraphics[width=0.5\textwidth]{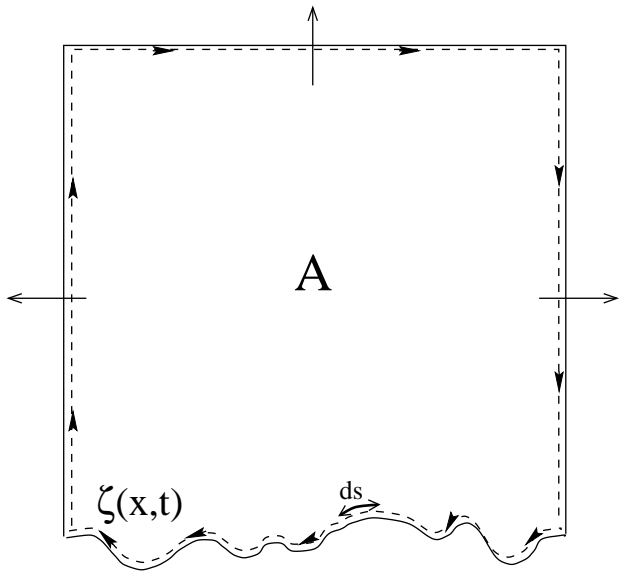}
\vspace{0.5cm}
\caption{Integration domain, $A$, and its boundary (solid line) used in Green
theorem. The infinitesimal arc length along the moving boundary $\zeta(x,t)$ is given by $ds$.}
\label{green_theorem}
\end{figure}

This technique has been used in other similar problems, such as solidification~\cite{Karma}
or epitaxial growth on vicinal surfaces~\cite{Pierre}, and is based on the use of the
Green theorem \cite{haberman} to transform an integral extended over a certain domain
(in our case, say, the region
between the electrodes (see Fig.\ \ref{green_theorem}) to an integral that is evaluated
precisely at the moving boundary (the aggregate surface).

Let us consider that the distance separating the electrodes and the lateral
size of the system are both infinite, so that the only part of the dashed
line in Fig.\ \ref{green_theorem} whose contribution is non vanishing is
the aggregate surface.
The Green function related to Eq.\ (\ref{n_difusion_eq}) is the solution
of
\begin{equation}
\left(\frac{\partial}{\partial t^\prime}+D\nabla^\prime{}^2
-V\frac{\partial}{\partial z^\prime}\right)G({\bf r}-{\bf
r}^\prime,t-t^\prime)= -\delta({\bf r}-{\bf r}^\prime)\delta(t-t^\prime),
\label{greendef}
\end{equation}
where we have made a change of coordinates to a frame of reference moving with the
average growth velocity, $V$. To evaluate $G$, we use its Fourier transform
\begin{equation}
G({\bf r}-{\bf r}^\prime,t-t^\prime)=\frac{1}{(2\pi)^3}\int d\omega\, e^{i\omega(t-t^\prime)}
\int d^2{\bf k}\, e^{i{\bf k}\cdot({\bf r}-{\bf r}^\prime)}G_{{\bf k}\omega},
\label{fouriergreen}
\end{equation}
where ${\bf r}=x{\bf \hat x}+z{\bf \hat z}$
and ${\bf k}=k_x{\bf \hat x}+k_z{\bf \hat z}$.
Hence,  Eq.\ (\ref{greendef}) becomes
\begin{equation}
G_{{\bf k}\omega}=\frac{1}{Dk^2+i\omega-i V k_z},
\end{equation}
with  $k^2=k_x^2+k_z^2$. Therefore, integrating Eq.\ (\ref{fouriergreen})
we find
\begin{equation}
\label{green}
G({\bf r}-{\bf r}^\prime,t-t^\prime)=\frac{\Theta(t-t^\prime)}{4\pi D(t-t^\prime)}
\exp\left[-\frac{(x-x^\prime)^2}{4D(t-t^\prime)}-
\frac{\big(z-z^\prime+V(t-t^\prime)\big)^2}{4D(t-t^\prime)}\right],
\end{equation}
$\Theta(t-t^\prime)$ being the Heaviside step function. In the following,
we will use for brevity $\tau\equiv t-t^\prime$.
It can be straightforwardly seen that the following relations are satisfied:
\begin{eqnarray}
\label{limeps}
\lim_{\tau \rightarrow 0^+} G({\bf r}-{\bf r}^\prime;\tau)&=&\delta({\bf r}-{\bf r}^\prime),\\
\label{liminfty}
\lim_{\tau\rightarrow-\infty}G({\bf r}-{\bf r}^\prime;\tau)&=&0,\\
\label{semiint}
\int_{-\infty}^{t}dt^\prime\int_{-\infty}^{\infty}dx^\prime G&=&\left\{\begin{array}{ll}
\exp[-(z-z^\prime)V/D]/V&\textrm{si $z>z^\prime$},\\
1/V&\textrm{si $z<z^\prime$},\end{array}\right.
\end{eqnarray}
We can now rewrite Eq.\  (\ref{n_difusion_eq}) in terms of the
variables ${\bf r^\prime}$ and $t^\prime$,
\begin{equation}
\left(\frac{\partial}{\partial t^\prime}-D\nabla^\prime{}^2
-V\frac{\partial}{\partial z^\prime}\right)c^\prime=-\nabla^\prime\cdot{\bf
q^\prime} \label{difprima}
\end{equation}
Adding Eq.\ (\ref{difprima}) multiplied by $G({\bf r}-{\bf r}^\prime,t-t^\prime)$ to
Eq.\ (\ref{greendef}) multiplied by $c^\prime\equiv c(x^\prime,z^\prime;t^\prime)$, and
integrating $t^\prime$ in  $(-\infty,t-\epsilon]$ and ${\bf r}^\prime$ in the set
$A=(-\infty,\infty)\times[\zeta(x^\prime,t^\prime),\infty)$, see Fig.\ \ref{green_theorem},
\begin{eqnarray}
\int_{-\infty}^{t-\epsilon}dt^\prime\int_{A}d^2{\bf r}^\prime
\frac{\partial}{\partial t^\prime}(c^\prime G)
-V\int_{-\infty}^{t-\epsilon}dt^\prime\int_{A}d^2{\bf
r}^\prime \frac{\partial}{\partial z^\prime}(c^\prime G)+\nonumber\\
D\int_{-\infty}^{t-\epsilon}dt^\prime\int_{A}d^2{\bf r}^\prime
\Big(c^\prime\nabla^\prime{}^2G-G\nabla^\prime{}^2c^\prime\Big)=
-\int_{-\infty}^{t-\epsilon}dt^\prime\int_{A}d^2{\bf r}^\prime
G\nabla^\prime\cdot{\bf q}^\prime
\label{partida}
\end{eqnarray}
The first term on the left hand side can be easily evaluated with the use of
Eqs.\ (\ref{limeps}) and (\ref{liminfty}). Thus,
\begin{equation}
\lim_{\epsilon\rightarrow 0}\int_{-\infty}^{t-\epsilon}dt^\prime\int_{A}d^2{\bf r}^\prime
\frac{\partial}{\partial t^\prime}(c^\prime G)=c({\bf r};t)+
\int_{-\infty}^{t}dt^\prime\int_{-\infty}^{\infty}dx^\prime
\frac{\partial\zeta^\prime}{\partial t^\prime}\Big[c^\prime G\Big]_{z^\prime=\zeta^\prime},
\label{I1}
\end{equation}
where $\zeta^\prime$ stand for $\zeta(x^\prime,t^\prime)$.
Analogously, we can integrate the second term on the left hand side of
Eq.\ (\ref{partida}), using (\ref{semiint})
\begin{equation}
\lim_{\epsilon\rightarrow
0}-V \int_{-\infty}^{t-\epsilon}dt^\prime\int_{A}d^2{\bf
r}^\prime \frac{\partial}{\partial z^\prime}(c^\prime G)=-c_a
+V\int_{-\infty}^{t}dt^\prime\int_{-\infty}^{\infty}dx^\prime \Big[c^\prime
G\Big]_{z^\prime=\zeta^\prime}.
\end{equation}
Finally, using the identity
$c^\prime\nabla^\prime{}^2G-G\nabla^\prime{}^2c^\prime
=\nabla^\prime \cdot(c^\prime\nabla^\prime{} G-G\nabla^\prime{} c^\prime)$, and
applying the Green theorem on the domain $A$, we get
\begin{equation}
\lim_{\epsilon\rightarrow 0}D\int_{-\infty}^{t-\epsilon}dt^\prime\int_{A}d^2{\bf r}^\prime
\left(c^\prime\nabla^\prime{}^2G-G\nabla^\prime{}^2c^\prime\right)=-D\int_{-\infty}^{t}dt^\prime
\int_{\zeta^\prime}ds^\prime\left(c^\prime\frac{\partial G}{\partial n^\prime}-
G\frac{\partial c^\prime}{\partial n^\prime}\right),
\end{equation}
$ds^\prime$ being the arc length, see Fig.\ \ref{green_theorem}, from which
we obtain the integro-differential equation
\begin{eqnarray}
c({\bf r},t)&=&c_a-\int_{-\infty}^{t}dt^\prime\Bigg[
\int_{-\infty}^{\infty}dx^\prime\left( V+ \frac{\partial
\zeta^\prime}{\partial t^\prime}\right)c^\prime G-\nonumber \\
&&D\int_{\zeta^\prime}ds^\prime\left( c^\prime\frac{\partial
G^\prime}{\partial n^\prime} -G\frac{\partial (c_0^\prime+c_1^\prime)}{\partial
n^\prime}\right)\Bigg]_{z^\prime=\zeta^\prime}-\sigma({\bf r},t),
\label{bulk_eq}
\end{eqnarray}
where
\begin{equation}
\sigma({\bf r},t)=\int_{-\infty}^{t}dt^\prime\int_{-\infty}^{\infty}dx^\prime
\int_{\zeta^\prime}^{\infty}dz^\prime G \nabla^\prime{}\cdot{\bf q}^\prime
\end{equation}
is a term related to the diffusion noise.

As we seek to determine $c$ everywhere, we need to know its  value at the
boundary. Considering the limit in which $\bf r$ belongs to that
boundary (hereafter, we will denote it as ${\bf r}_b$),
the term  $\partial G/\partial n^\prime$ in Eq.\ (\ref{bulk_eq}) is singular.
Fortunately, this singularity is integrable, as a result of which we obtain an additional term as~\cite{Pierre}
\begin{equation}
\int_{{\bf r}\notin \zeta}c^\prime\frac{\partial G}{\partial
n^\prime}\rightarrow \frac{c}{2D}+ \int_{{\bf r}_b\in\zeta}
c^\prime\frac{\partial G}{\partial n^\prime} .
\end{equation}
This leads to Eq.\ (\ref{main_eq}) of the main text, where we have omitted the subindex $b$
for notational simplicity.

\section{Zeroth order calculation}

\label{zeroth_order}

Writing $c=c_0+c_1$ in Eq.\ \eqref{main_eq} we get
\begin{equation}
\frac{c_0}{2}+\frac{c_1}{2}=c_a-\int_{-\infty}^{t}dt^\prime
\int_{-\infty}^{\infty}dx^\prime\Bigg[V\left(c_0^\prime G+c_1^\prime
G\right)+
\frac{\partial \zeta^\prime}{\partial t^\prime}c_0^\prime G
-D\Bigg( (c_0^\prime+c_1^\prime)\frac{\partial
G}{\partial n^\prime}-
 G \frac{\partial (c_0^\prime+c_1^\prime)}{\partial n^\prime} \Bigg)
\Bigg]_{z^\prime=\zeta^\prime}-{\sigma}({\bf r},t) .
\label{linear_main_eq}
\end{equation}
Note that, at this order, $ds^\prime=\sqrt{1+(\partial_x\zeta)^2}dx^\prime\simeq dx^\prime$.
We also linearly expand $G$, so that
\begin{equation}
\label{linear_green}
G({\bf r}-{\bf r}^\prime,t-t^\prime)\simeq\Big(1
-V\frac{\zeta-\zeta^\prime}{2D}\Big)G^0,
\end{equation}
where
\begin{equation}
G^0=\frac{\Theta(\tau)}{4\pi D\tau}\exp\Big[-\frac{(x-x^\prime)^2}{4D\tau}
-\frac{V^2\tau}{4D} \Big].
\end{equation}
In order to determine the concentration at the boundary, we must write
our main equation in terms of $c_0$ and $c_1$. Then, with the use of the
boundary condition (\ref{n_boundary_cond}) we find
\begin{equation}
D\frac{\partial (c_0+c_1)}{\partial n}=k_D(c_0+c_1-c_{eq}^0
+\Gamma \partial^2_x\zeta+\chi)+{\bf q}\cdot{\bf n}.
\label{linear_frontera}
\end{equation}
Finally, putting Eqs.\ (\ref{linear_green}) and (\ref{linear_frontera}) into
(\ref{linear_main_eq}) we find
\begin{equation*}
\frac{c_0}{2}+\frac{c_1}{2}=c_a+\int\!\!\!\int dt^\prime dx^\prime
\Bigg[\bigg(-c_0
\frac{\partial\zeta^\prime}{\partial t^\prime}
-(k_D+V/2)(c_0^\prime + c_1^\prime)+
k_DV\frac{c_0^\prime}{2D}(\zeta-\zeta^\prime)+
k_Dc_{eq}^0\Big(1-\frac{\zeta-\zeta^\prime}{2D}V\Big)-
\end{equation*}
\begin{equation}
- \Gamma k_D\zeta_{x^\prime x^\prime}^\prime
+c_0\frac{\zeta-\zeta^\prime}{2}\bigg(\frac{V^2}{2D}+\frac{1}{\tau}\bigg) -
 c_0\frac{(x-x^\prime)\partial_{x^\prime}\zeta^\prime}{2\tau}
\bigg)G^0\Bigg]_{z^\prime=\zeta^\prime}-{\tilde\sigma}({\bf r},t),
\label{linear_main2}
\end{equation}
with a new noise term
\begin{equation}
{\tilde \sigma}({\bf r},t)=\int_{-\infty}^{t}dt^\prime
\int_{-\infty}^{\infty}dx^\prime \bigg[ (k_D\chi^\prime +{\bf q}^\prime\cdot{\bf n^\prime})G^0 +
 \int_{\zeta^\prime}^\infty dz^\prime G^0\nabla^\prime\cdot{\bf q}^\prime\bigg].
\end{equation}
Despite the apparent complexity of these new equations,
Eq.\ (\ref{linear_main_eq}) is linear so that Fourier transforming it we get the
following algebraic equation which relates all the zeroth order terms
\begin{equation}
\frac{c_{k\omega}^0}{2}=\left(c_a+\frac{k_Dc_{eq}^0}{V}\right)\delta(k)\delta(\omega)-
\left(\frac{V}{2}+k_D\right)c_{k\omega}^0G_{k\omega}^0,
\label{c1kw}
\end{equation}
$c_{k\omega}^0$ being the Fourier transformation of $c_0$ and
\begin{equation}
G_{k\omega}^0=\Big[4D\omega i+4D^2k^2+V^2 \Big]^{-1/2},
\label{fourier_green}
\end{equation}
which can be easily inverted yielding Eq. (\ref{c0_b}).

\section{First order calculation}
\label{first_order}

From the results obtained in Apps.\ \ref{green_ap} and \ref{zeroth_order}
we can find an evolution equation which relates $c_{k\omega}^1$ and
$\zeta_{k\omega}$. Thus,
\begin{equation*}
\frac{c^1_{k\omega}}{2}=\Bigg[-i\omega c_0-(k_D+V/2)\frac{c^1_{k\omega}}{\zeta_{k\omega}}+
\frac{1}{2D}\,\big(k_Dc_0-
k_Dc^0_{eq}\big)\bigg(\frac{1}{G^0_{k\omega}}- V\bigg)+ \Gamma k_Dk^2+
\end{equation*}
\begin{equation}
+ \frac{c_0}{4D}\,\bigg(\frac{1}{(G^0_{k\omega})^2}-V^2\bigg)-
\frac{c_0}{2}Dk^2\Bigg]G^0_{k\omega}\zeta_{k\omega}-
 G^0_{k\omega}{\tilde\sigma}_{k\omega} ,
\label{matrix1_pre}
\end{equation}
${\tilde\sigma}_{k\omega}$ being the Fourier transformation of
${\tilde\sigma}({\bf r},t)$, hence
\begin{equation}
c^1_{k\omega}\left(\frac{1}{2G^0_{k\omega}}+\frac{V}{2}+k_D \right)=
\Bigg[\frac{V}{\Omega D}\left(\frac{1}{2G^0_{k\omega}}-\frac{V}{2}
\right)+k_D\Gamma k^2\Bigg]\zeta_{k\omega}-{\tilde\sigma}_{k\omega}.
\label{matrix1}
\end{equation}
This equation has two unknowns; hence, in order to solve it, we also need to
expand Eq.\ (\ref{n_velocidad}) in powers of $c_1$ and $\zeta$. Thus,
\begin{equation}
\label{matrix2}
c_{k\omega}^1=\bigg(\frac{i\omega}{\Omega k_D}+\Gamma k^2+
\frac{B}{k_D}k^4
\bigg)\zeta_{k\omega}-\chi_{k\omega}+\frac{ik}{k_D}p_{k\omega}.
\end{equation}
Combining both Eqs.\ (\ref{matrix1}) and (\ref{matrix2}), we find
\begin{equation}
{\mathcal
T}_{k\omega}\zeta_{k\omega}=\beta_{k\omega},
\label{transfer_eq}
\end{equation}
with
\begin{equation}
{\mathcal T}_{k\omega}=\bigg(\frac{1}{2G_{k\omega}^0}+\frac{V}{2}+k_D\bigg)
\bigg(\frac{i\omega}{\Omega k_D}+\Gamma k^2+\frac{B}{k_D}k^4\bigg)
-
 \frac{V}{\Omega D}\bigg( \frac{1}{2G_{k\omega}^0}-\frac{V}{2}\bigg)-
k_D\Gamma k^2.
\label{transfer}
\end{equation}
The new noise term $\beta_{k\omega}$ is the projection of all the noise terms onto the
boundary, and is given by
\begin{eqnarray}
\beta_{k\omega}&=&\bigg(\frac{1}{2G_{k\omega}^0}+\frac{V}{2}+k_D\bigg)\left(
\chi_{k\omega}-\frac{ik}{k_D}p_{k\omega}\right)-{\tilde\sigma}_{k\omega}=
\nonumber \\
&=&D\Lambda^{(+)}_{k\omega}\chi_{k\omega}-
\frac{ik}{k_D}p_{k\omega}(D\Lambda^{(+)}_{k\omega}+k_D)+
\int _0^\infty dz^\prime\left(ikq^x_{k\omega}+
q^z_{k\omega}\Lambda^{(-)}_{k\omega}
\right)\exp(-\Lambda^{(-)}_{k\omega}z^\prime) ,
\end{eqnarray}
where
\begin{equation}
\Lambda^{(\pm)}_{k\omega}=\frac{1}{2DG^0_{k\omega}}\pm\frac{V}{2D}.
\end{equation}
These equations allow us to calculate the noise correlations (in Fourier space)
\begin{equation*}
\langle\beta_{k\omega}\beta_{k^\prime\omega^\prime}\rangle =
\left[ \frac{2D^2c_0}{k_D}|\Lambda^{(+)}_{k\omega}|^2+
2Dc_0\left(\frac{k^2+|\Lambda^{(-)}_{k\omega}|^2}{2\mathrm{Re}(\Lambda_{k\omega}^{(-)})}\right)+
2D_s\nu\frac{k^2}{k_D^2}\Big(D^2|\Lambda^{(+)}_{k\omega}|^2
+k_D^2+
 2Dk_D\mathrm{Re}(\Lambda^{(+)}_{k\omega}) \Big)\right]\times
\end{equation*}
\begin{equation}
\label{betabeta}
\times \delta(k+k^\prime)\delta(\omega+\omega^\prime).
\end{equation}
Note that, in principle, Eq.\ (\ref{transfer_eq}) provides us with all the information of the
system at linear order, namely, the power spectrum of $\zeta$ (by evaluating
${\mathcal T}_{k \omega}^{-1}$ and then integrating out the temporal frequency $\omega$)
or the height-height correlations (integrating out the spatial frequency $k$).

In order to gain insight about the implications of this expansion in $\zeta$, we write
Eq.\ (\ref{transfer_eq}) as the Fourier transformation of a Langevin equation for the interface height,
\begin{equation}
[i\omega-\omega_k]\zeta_{k\omega}=\eta_{k\omega},
\label{reescritura}
\end{equation}
$\omega_k$ being a function of $k$ which we must specify from ${\mathcal T}_{k \omega}$,
and $\eta_{k\omega}$ being a noise term related with
$\beta_{k \omega}$, that is also delta-correlated,
\begin{equation}
\langle \eta_{k\omega}\eta_{k^\prime\omega^\prime}\rangle=\Pi(k)\delta(k+
k^\prime)\delta(\omega+\omega^\prime),
\end{equation}
where $\Pi$ depends, in general, on $k$ and provides the magnitude of the noise for each
Fourier mode, see Sect.\ \ref{nee}. Finally, by Fourier transforming the time frequency,
we find the linear stochastic partial differential equation (\ref{pre_evol_eq}).

\bibliography{nicoli}

\begin{thebibliography}{10}
\bibitem{noteaboutspecies}
Sometimes the system contains also other noncharged molecules that affect only
  the effective mass transport ({\em i.e.}, the diffusivities) but not the main
  evolution equations.
\bibitem{tercergrado}
We can find those zeros by writing Eq.\ (\ref{transfer}) as a third order
  polynomial in a new variable $\Lambda_k\equiv
  \frac{1}{2DG^0_{k\omega}}+\frac{V}{2D}$, which is simpler to deal with.
\end{thebibliography}


\end{document}